\documentclass[12pt]{article}

\usepackage{epsf,epsfig}
\usepackage{axodraw}

\usepackage{a4wide}
\def\slash#1{#1 \hskip-0.45em /}
\def\Slash#1{#1 \hskip-0.59em /}

\def\be{\begin{equation}}
\def\ee{\end{equation}}
\def\beq{\begin{eqnarray}}
\def\eeq{\end{eqnarray}}

\def\eps{\epsilon}

\def\np{n_+}
\def\nm{n_-}

\def\Cl2{\mbox{Cl}_2}
\def\eps{\epsilon}

\def\slash#1{#1 \hskip-0.45em /}
\def\Slash#1{#1 \hskip-0.59em /}

\def\cO#1{{\cal O}\left( {#1} \right)}
\def\cD{{\cal D}}
\def\eqref#1{(\ref{#1})}

\def\nn{\nonumber}

\newcommand{\lnM}{\ln\frac{\hat M^2}{\mu^2}}
\newcommand{\lnsqM}{\ln^2\frac{\hat M^2}{\mu^2}}

\def\cL{{\cal{L}}}
\def\hM{\hat M}
\def\Mmue{\left(\frac{\hat M^2}{\mu^2}\right)^{-\epsilon}}
\def\Mmutwoe{\left(\frac{\hat M^2}{\mu^2}\right)^{-2\epsilon}}

\def\eps{\epsilon}

\def\MSbar{\overline{\rm MS}}

\begin{document}
\thispagestyle{empty}

\begin{flushright}
  PITHA 03/14\\
  IPPP/03/83\\
  FERMILAB-Pub-03/410-T\\
  hep-ph/0401002\\
  December 31, 2003 
\end{flushright}

\vspace{\baselineskip}

\begin{center}
\vspace{0.5\baselineskip}
\textbf{\Large 
Effective theory calculation of resonant \\[0.2cm]
high-energy scattering}\\
\vspace{3\baselineskip}
{\sc M.~Beneke$^a$, A.P.~Chapovsky$^a$, A.~Signer$^b$ and 
G.~Zanderighi$^c$}\\
\vspace{0.7cm}
{\sl ${}^a$Institut f\"ur Theoretische Physik E, RWTH Aachen,\\
D--52056 Aachen, Germany\\
\vspace{0.3cm}
${}^b$IPPP, Department of Physics, University of Durham, \\
Durham DH1 3LE, England\\
\vspace{0.3cm}
${}^c$Fermi National Accelerator Laboratory, Batavia, 
        IL 60510-500, USA}
\vspace{3\baselineskip}

\vspace*{0.2cm}
\textbf{Abstract}\\
\vspace{1\baselineskip}
\parbox{0.9\textwidth}{
Tests of the Standard Model and its hypothetical extensions require
precise theoretical predictions for processes involving massive,
unstable particles. It is well-known that ordinary weak-coupling
perturbation theory breaks down due to intermediate singular
propagators. Various pragmatic approaches have been developed to deal
with this difficulty. In this paper we construct an effective field
theory for resonant processes utilizing the hierarchy of scales
between the mass of the unstable particle, $M$, and its width,
$\Gamma$. The effective theory allows calculations to be
systematically arranged into a series in $g^2$ and $\Gamma/M$, and
preserves gauge invariance in every step. We demonstrate the
applicability of this method by calculating explicitly the inclusive
line shape of a scalar resonance in an abelian gauge-Yukawa model at
next-to-leading order in $\Gamma/M$ and the weak couplings. We also
discuss the extension to next-to-next-to-leading order and compute an
interesting subset of these corrections.}
\end{center}

\newpage
\setcounter{page}{1}

\newpage

\section{Introduction}
\label{sec:intro}

An important part of recent and future high-energy physics
experiments belongs to the detailed investigation 
of the production of heavy unstable particles, such as $W$ and $Z$
bosons, top quarks, Higgs bosons or perhaps other new 
particles. From single resonant or pair production the masses 
and couplings of the particles can be determined with high 
precision, provided theoretical calculations are equally 
precise. Since the decay widths of the particles are often 
non-negligible, this includes a consistent treatment of 
finite-width effects beyond the narrow-width approximation. 

Several higher-order calculations involving unstable particles have
been performed in recent years, in particular for the line shape 
of the $Z$ boson \cite{lep1},
$W$-pair production \cite{ww-calc}, and 
$t\bar{t}$-production \cite{tt-calc}.
In these calculations the finite width of the particles has 
been treated in a variety of often pragmatic approaches. While 
this may be adequate for the present, it is certainly desirable 
to formulate a theoretical framework that would allow for  
systematic improvements of the accuracy of such calculations. 
Moreover, future precision experiments require that such a framework 
be developed. 

The difficulty with unstable particles is that they cause 
singularities in propagators, if the scattering amplitudes 
are constructed according to the rules of weak-coupling perturbation 
theory. A well-known remedy of the singularity is 
resummation of self-energy corrections 
to the propagator, which results in the substitution
\begin{equation}
\frac{1}{p^2-M^2}\,\to\,\frac{1}{p^2-M^2-\Pi(p^2)}.
\end{equation} 
The self-energy has an imaginary part of order $M^2 g^2 \sim 
M\Gamma$, where $\Gamma$ is the on-shell decay width of the 
resonance, rendering the propagator large but finite. ``Dyson 
resummation'' sums a subset of terms of order 
$(g^2 M^2/[p^2-M^2+i M\Gamma])^n\sim 1$ (near resonance) 
to all orders in the expansion in the coupling $g^2$. This procedure 
raises the question of how to identify 
all terms (and only these terms) required to achieve a specified 
accuracy in $g^2$ and $\Gamma/M$. The failure to address this 
question may lead to a lack of gauge invariance and unitarity of 
the resummed amplitude, since these properties are guaranteed only 
order-by-order in perturbation theory, and for the exact 
amplitude. 

Many of the current approaches to unstable particles take the
restoration of gauge invariance as their starting point. One example 
is the fermion-loop scheme $\cite{fls}$, which is based on the 
observation that the dominant contribution to the width of the $W$ and
$Z$ gauge bosons comes from fermion loops. The prescription reads 
to include the fermion-loop corrections in propagators and vertices, 
so that gauge invariance is maintained, since all fixed-order $g^2 N_f$
terms (where $N_f$ is the number of fermion flavours) are included. 
Besides the restriction to gauge bosons the disadvantage of this
scheme is that one-loop vertices must be computed even for a
leading-order approximation. Another scheme 
\cite{nonlocal} constructs a gauge-invariant non-local effective action,
which can be matched onto the two-point functions of the 
underlying theory. The gauge Ward identities are satisfied by 
construction in this scheme. Both schemes have been 
implemented in four-fermion production mediated by $W$-pair production
solving not only the 
gauge-invariance problem but also capturing some sophisticated
features such as running of the couplings. However, it is not 
evident how to extend them to a systematic approximation of the
scattering amplitude in powers of $g^2$ and $\Gamma/M$. 

Other approaches exploit the presence of two different
momentum scales in the production and decay of weakly interacting 
unstable particles: $M$, the mass of the unstable particle, 
and its width, $\Gamma\ll M$, both related to the location of the 
resonance pole. In the pole scheme \cite{pole-scheme} the 
scattering amplitude is approximated by an expansion in 
$\Gamma/M$ around the poles in the complex plane. The coefficients 
of this expansion must be gauge-invariant in every order. 
In pair production of unstable particles this approximation is
referred to as ``double-pole approximation'' \cite{ww-calc}.
It has been used in the complete next-to-leading order 
(NLO) calculation for $W$-pair production \cite{ww-calc}, 
where in one-loop radiative corrections the leading term in the 
$\Gamma/M$ expansion is sufficient.

The pole approximation can be considered as the 
first step towards a systematic approximation scheme to the 
scattering amplitude based on the separation of scales. 
Within the (double) pole approximation the process naturally consists 
of production and decay subprocesses, connected by the intermediate 
resonance(s). One-loop corrections separate into
two subsets: {\it factorizable corrections}\/ to the hard
production/decay subprocesses, and {\it non-factorizable
  corrections}, accounting for interference between
different hard subprocesses.
This separation does not occur at the level of individual Feynman 
diagrams. For example, the self-energy correction to the 
unstable particle contains both, factorizable and 
non-factorizable pieces, and the two have to be separated carefully 
to avoid double counting. This is expected to be complicated 
for multi-loop calculations, or for a calculation of $\Gamma/M$
suppressed contributions, which lie beyond the pole approximation.
One-loop non-factorizable corrections have been extensively studied 
in the double-pole approximation. A few important theorems have 
been proved \cite{nf-theorems}, and explicit calculations
have been performed \cite{nf-calc}. 
We also mention an approach 
that constructs the expansion in $g^2$ of the resonant propagator 
squared in the distribution sense \cite{Tkachov:1999qb}, also
addressing the issues of gauge invariance and systematic expansions. 

In \cite{effective} it was suggested to use the scale 
hierarchy $\Gamma\ll M$ for constructing an
effective field theory, from which hard modes with momenta 
of order $M$ or larger are removed. 
The effect of the hard modes is included into the
coefficients of the effective Lagrangian and
corresponds to the factorizable corrections. Non-factorizable
effects are reproduced by the dynamical modes of the  
effective theory. This provides a more precise definition 
of ``factorizable'' and ``non-factorizable'' that generalizes 
to higher loop orders. In \cite{effective} 
it was shown that this approach is equivalent
to the double-pole approximation within the accuracy of the latter 
(i.e.~one-loop, leading order in $\Gamma/M$). However, the idea 
should work beyond these approximations.

In this paper we pursue this idea and develop the effective 
field theory approach to unstable particle production from a 
systematic point of view. We take the attitude that the 
scale hierarchy $\Gamma \ll M$ is {\em the} characteristic feature of 
the process, and that all other issues such as resummation and gauge
invariance will be an automatic consequence of any theory that
formulates the rules for an expansion in $g^2$ and $\Gamma/M$
correctly. We identify the factorization properties of 
the process to any order in $\Gamma/M$, and define the effective 
production and decay vertices and the effective Lagrangian 
in dimensional regularization. The matching corrections are computed 
by expansion of Feynman integrals in momentum 
regions \cite{Beneke:1998zp}. In this way we identify all
contributions to the scattering amplitude at a given order in 
$g^2$ and $\Gamma/M$ (and only these). We believe that this approach 
solves the conceptual difficulties that have so far been 
associated with the perturbative treatment of unstable particles 
in principle; however, for complex scattering processes, the
implementation in higher orders than NLO still requires difficult 
calculations. 

Here we shall describe our approach for the case of 
inclusive production of a single charged scalar resonance 
in fermion-fermion scattering in an abelian gauge model. The specific
set-up of the toy model, and the restriction to inclusive scattering, 
imply that a minimal set of effective fields needs to be 
introduced, and only the forward scattering amplitude needs to be 
considered. This allows us to concentrate on the essential features 
of the approach, such as the identification of momentum modes, 
factorization of the scattering amplitude, construction of the 
effective Lagrangian and matching, and the calculation of the
scattering amplitude in the effective theory. We compute explicitly the 
inclusive line shape of the scalar particle in this toy model, and 
discuss some features at the next-to-next-to-leading order related 
to gauge invariance and $g^2\Gamma/M$ corrections. The generalization 
of the method to final states with detected particles or jets, 
to non-abelian gauge bosons, and to pair production appears not to 
encounter major new conceptual issues, but requires more technical 
work, which we hope to complete in the future. A brief exposition 
of the effective theory approach discussed in the present 
paper has been given in \cite{letter}.

The outline of the paper is as follows. In Section~\ref{sec:method} we
define our toy theory and proceed to discuss in a non-technical manner
the ingredients of the effective field theory method: the momentum
scales, the presence of soft and collinear modes, the effective
interactions and the expression for the forward scattering amplitude
in the effective theory. We outline a hierarchy of effective theories,
in which collinear fields are either kept or integrated out. The
section concludes with a leading-order calculation of the line shape
and a discussion of scheme-dependence and matching to the
off-resonance cross section. The real virtue of the effective theory
approach becomes only apparent when one goes beyond leading order. A
complete next-to-leading order calculation is carried out in
Section~\ref{sec:nlo}, where we compute the two-loop matching of the
two-point interaction, and the one-loop matching of the production
vertex, required at NLO. The NLO result is completed with the
calculation of the forward scattering amplitude in the effective
theory. At the end of Section~\ref{sec:nlo} we perform a numerical
comparison of the leading-order and next-to-leading-order line
shape. The formalism is sufficiently general to allow a systematic
extension to next-to-next-to-leading order. While there is no point of
doing this calculation in our toy model, we outline the principles of
such a calculation in Section~\ref{sec:beyondnlo}. Then, as an
example of a  next-to-next-to-leading order contribution, 
we compute the one-loop short-distance coefficient of the 
$\Gamma/M$-suppressed production-decay operators and illustrate how a
gauge-independent result is obtained automatically from a combination
of self-energy, vertex, and box diagrams. 
We conclude in Section~\ref{sec:conclude}.  The
renormalization constants of the toy theory are collected in an
Appendix.

\section{Outline of the method}
\label{sec:method}

In this section we explain the essential features of our approach. 
We first set up the toy field theory and scattering process that we 
consider in this paper. We then discuss the short- and long-distance 
scales in resonant scattering, the corresponding momentum modes, and 
derive a representation of the scattering amplitude, in which the 
different scales are factorized, allowing for an expansion in 
$\Gamma/M$. We end this section by giving the two-to-two forward 
scattering amplitude at leading order, and obtain from this the 
leading-order line shape of the resonance.

\subsection{Definition of the model}
\label{sec:modeldef}

We shall consider the resonant production of a massive scalar particle 
in the scattering of two massless Dirac fermions. The scalar and one 
of the fermions (the ``electron'') are assumed to be charged 
under an abelian gauge 
symmetry, the other fermion (the ``neutrino'') is neutral. The charges 
are supposed to be equal, so as to allow a scalar-electron-neutrino 
Yukawa coupling. The theory is much simpler than the electroweak 
Standard Model, but we find that all conceptual issues related to 
the treatment of the scalar resonance can be addressed in this toy 
model. 

We thus assume the Lagrangian
\begin{eqnarray}
\label{model}
{\cal L} &=& (D_\mu\phi)^\dagger D^\mu\phi - \hat M^2 \phi^\dagger\phi + 
 \bar\psi i \!\not\!\!D\psi + \bar\chi i\slash{\partial}\chi - 
 \frac{1}{4} F^{\mu\nu}F_{\mu\nu}-\frac{1}{2\xi} \,(\partial_\mu
 A^\mu)^2 
 \nonumber\\ 
 && + \, y\phi\bar\psi\chi + y^* \phi^\dagger \bar\chi\psi 
-\frac{\lambda}{4}(\phi^\dagger \phi)^2+ {\cal L}_{\rm ct},
\end{eqnarray}
where ${\cal L}_{\rm ct}$ denotes the counterterm Lagrangian, and 
the scalar self-interaction is included to make the model 
renormalizable. Fields 
and parameters are renormalized in the $\overline{\rm MS}$ scheme. 
The letter $\hat M$ is used to distinguish the 
$\overline{\rm MS}$ mass (at scale $\mu$) from the pole mass $M$ 
defined below. Explicit expressions for the counterterm Lagrangian 
and the renormalization constants are given in the Appendix. 
The gauge coupling $g$ and Yukawa
coupling $y$ are assumed to be of the same order. We define 
$\alpha_g=g^2/(4\pi)$, $\alpha_y=(y y^*)/(4 \pi)$, and use $\alpha$ to 
refer to them summarily. With no arguments the couplings 
are evaluated at scale $\mu$. In general, we cannot set the
scalar self-coupling $\lambda$ to zero, but we would like it to be
small (for simplicity of the model). Without fine-tuning it is
consistent to assume that 
$\alpha_\lambda/(4\pi)\equiv\lambda/(16\pi^2)$ 
is of order $\alpha^2/(4\pi)^2$, since the leading counterterm is of
this order. We shall assume this counting in the following. 

We shall consider the totally inclusive cross section in 
electron-neutrino scattering, 
\be 
\bar\nu(q) + e^-(p)\to X,
\label{process}
\ee
as a function of the center-of-mass energy squared 
$s=(q+p)^2$ in the vicinity of $s\approx M^2$, where we expect an 
enhancement of the cross section due to the resonant production of 
the charged scalar. By vicinity we mean that $s-M^2\sim 
M \Gamma \sim M^2 \alpha 
\ll M^2$. Because the electron is massless, the total cross 
section is not infrared-safe. The initial-state collinear 
singularity should be factorized into the electron distribution 
function. This being understood, we will usually quote the ``partonic'' 
cross section with the singularity regularized in dimensional 
regularization and subtracted minimally. Note that strictly speaking 
the electron distribution function is not defined if the electron is
truly massless as assumed in (\ref{model}); 
however, we may always assume that the electron 
has a mass much smaller than any other scale in the scattering
process. We can then neglect the mass in the Lagrangian, keeping in 
mind that it must be reintroduced to regularize the distribution 
function. 

Our aim is to approximate this totally inclusive line shape of the 
scalar resonance in electron-neutrino scattering systematically 
in powers of $g^2$ and 
\be
\label{eq:deltadef}
\delta\equiv \frac{s-\hat M^2}{\hat M^2} \sim \frac{\Gamma}{M}.
\ee
This cannot be done with standard methods, since there are 
kinematic enhancements proportional to $\alpha \hat M^2/(s-\hat M^2)\sim 1$ 
at every order in the perturbation expansion in $\alpha$. This 
is the origin of the well-known need for resummation. 

\subsection{Effective Lagrangians, effective vertices and 
representation of the scattering amplitude}
\label{sec:2.2}

In space-time the resonance is produced in the collision with a
characteristic formation time of order $1/M$, lives a much longer time
of order $1/\Gamma$, and then decays, again within a short time of 
order $1/M$. We therefore expect some kind of factorization between
production, propagation and decay. This essentially classical
space-time picture is corrected, because quantum fluctuations exist at
all scales. Factorization persists quantum-mechanically, since only
long-wavelength fluctuations can resolve the details of production and
decay (separated by the long time interval $1/\Gamma$)
simultaneously. The resulting picture is shown in the left-hand graph 
of Figure~\ref{fig:skeleton} below. (The right-hand graph shows 
the featureless ``background processes'', in which
the final state of the collision is produced without 
the resonance.) Factorization in the presence of quantum
corrections can be implemented with the effective
Lagrangian technique.

\subsubsection{Effective Lagrangian for soft and collinear interactions}

To obtain the expansion of the line-shape in $\delta$, 
we construct the effective Lagrangian and effective vertices for 
the long-distance contributions to the process. ``Hard'' 
effects related to quantum fluctuations with 
momenta $k\sim M$ are included as coefficient
functions by matching 
the effective Lagrangian to the underlying theory. This can be 
done in ordinary weak-coupling perturbation theory, since hard 
propagators, being off-shell by an amount of order $M^2$, do not 
cause kinematic enhancements proportional to $1/\delta$. Each 
Feynman diagram is broken into a hard and other 
contributions similar to the non-relativistic expansion of 
Feynman integrals \cite{Beneke:1998zp}. The hard region corresponds to 
the Taylor expansion of the Feynman {\em integrand} in $\delta$
(counting the loop momentum $k\sim M$), 
and goes into the coefficient functions of the effective Lagrangian 
and effective vertices. The remaining regions correspond to momentum 
configurations, where propagators are near mass-shell. They 
must be reproduced by the diagrams in the effective theory. 

What are the momentum modes in the effective theory? Parameterize the
momentum of the scalar particle near resonance by $P=\hat M v+k$,
where the velocity vector $v^2=1$ and the residual momentum $k$ scales
as $M\delta\sim\Gamma$. Then $P^2-\hat M^2$ remains small, of order
$M\delta$, if the scalar interacts with a ``soft'' fluctuation with
momentum of order $M\delta$. The effective theory therefore contains
soft massless modes, and the heavy scalar near mass-shell. (To
emphasize the fact that the form of the effective theory is very much
the same for stable and unstable particles, we will also call the
heavy scalar near mass-shell a ``soft'' mode. In previous work
\cite{effective} the term ``resonant mode'' has been used.) 
Note that soft interactions do not change the velocity vector of the
scalar. We therefore define a field $\phi_v(x)$ with the rapid spatial
variation $e^{-i \hat M v\cdot x}$ removed to represent the heavy
scalar with momentum near $\hat M v$. The field $\phi_v(x)$ carries
only the residual slow variation in $x$ over distances of order
$1/\delta$. If the heavy particle was a stable heavy quark,
the effective Lagrangian that describes its interactions with soft
quark and gluon fluctuations would be the heavy quark effective
Lagrangian \cite{Eichten:1990zv}, which has become a standard tool in
heavy quark physics to separate the physics on the scales $M$ and
$\Lambda_{\rm QCD}$. The corresponding Lagrangian for an unstable
scalar is
\be
\label{heavyLO}
{\cal L}_{\rm HSET} = 2\hat M\,\phi_v^\dagger 
\left(iv\cdot D_s-\frac{\Delta}{2}\right)\phi_v
-\frac{1}{4}\,F_{s\mu\nu}
F_s^{\mu\nu}+\bar\psi_s i\!\not\!\!D_s \psi_s+
\bar\chi_s i\!\not\!\partial \chi_s
+\ldots. 
\ee
Here $D_s=\partial-i g A_s$ denotes the covariant derivative with a 
soft photon field, and we have kept the relativistic normalization with 
mass dimension 1 for the non-relativistic scalar field $\phi_v$, hence 
the factor $2\hat M$. Furthermore, $\psi_s$ ($\chi_s$) stands for the 
soft electron (neutrino) field. Only the leading power 
in the expansion in $1/\hat{M}$ has been given, but there is no 
difficulty in adding further terms. The leading-power Lagrangian 
contains a single short-distance matching coefficient, $\Delta$, 
which can be related to the resonance pole position $M^2-i M\Gamma$ 
as we discuss below (Section~\ref{sec:heavy}). 
In the pole renormalization scheme $\Delta$ 
equals $-i\,\Gamma$. Since the Lagrangian describes only effects related 
to the soft scale $M\delta$, all dimensionful quantities can be 
assigned a scaling power of $\delta$. (In such assignments, we set 
$M=1$. Dimension is restored by inserting the appropriate power of 
$M$.) Partial derivatives and the soft photon field count 
as $\delta$. The field $\phi_v$ as well as soft massless fermion fields 
count as $\delta^{3/2}$, because their position-space propagator is 
proportional to $1/x^3\sim\delta^3$. In particular, since 
$\Delta$ is of order $\Gamma\sim M\alpha\sim\delta$, and so is 
$iv\cdot D_s$, the 
decay width of the heavy scalar, or, more generally, the coefficient 
$\Delta$, is a leading-power effect in the Lagrangian, and must be 
included in the scalar propagator. This is the effective theory 
counterpart of the familiar self-energy resummation.

The effective theory is not yet complete, because the scalar is 
produced in the scattering of energetic ($E\sim M$) massless particles. 
To describe these ``collinear'' modes, we suppose that the 
electron moves with large momentum in the direction of $\vec{n}_-$, 
and introduce two reference light-like vectors, $n_\pm$, with 
$n_+^2=n_-^2=0$ and $\np\nm = 2$. A collinear momentum is decomposed 
as  
\begin{equation}
  p^\mu = (n_+p) \, \frac{n_-^\mu}{2} + p_\perp^\mu + (n_-p)
  \frac{n_+^\mu}{2},
\end{equation}
where $n_+ p \sim M$, $n_- p \sim M\delta$ and 
$p_\perp \sim M\delta^{1/2}$. The scaling of the small components 
is determined by the interaction of collinear modes with 
soft modes, which implies that a generic collinear fluctuation 
has an off-shellness of order $\hat M\delta$. The scaling of the 
transverse component is then fixed by the poles of collinear 
propagators. Collinear modes in an effective 
field theory framework have been discussed previously in 
$B$ meson decays into light energetic mesons \cite{Bauer:2000ew}. 
Our set-up is in fact very similar to 
this case, since the production of the scalar resonance represents 
the ``inverse'' kinematics of $B$ decay. The neutrino plays the 
role of the weak current, so that we consider the production 
of a heavy particle by scattering an energetic electron on  
this ``current''. The effective Lagrangian for the interactions 
of collinear modes and soft modes is, again at leading 
power in the expansion in $\delta$, 
\be
\label{scetLO}
{\cal L}_{\rm SCET} = 
\bar{\psi}_c \left(i n_- D +  i \Slash{D}_{\perp c}
\frac{1}{i n_+ D_{c}+i\epsilon}\, i\Slash{D}_{\perp c} \right)
\frac{\slash{n}_+}{2} \, \psi_c -\frac{1}{4}\,F_{c\mu\nu} F_c^{\mu\nu}
+ \ldots,
\ee
where $\psi_c$ denotes the collinear electron field, which 
satisfies $\not\!\!n_-\psi_c=0$ \cite{Bauer:2000yr,BCDF}. 
The covariant derivative $i D_c=i\partial
+ g A_c$ contains the collinear photon field. 
The interaction with the soft field appears only in 
$i\nm D=i\nm\partial + g \nm A_c+g\nm A_s$.
Again there is no difficulty in going to higher orders in the expansion; 
we discuss this below (Section~\ref{sec:scet}). The $\delta$ power 
counting is as follows: the components of derivatives on collinear 
fields and the collinear photon field scale as the corresponding 
collinear momentum components; the collinear fermion field 
scales as $\psi_c\sim \delta^{1/2}$ as seen from the 
propagator. Note that there 
do not exist collinear interactions with the heavy scalar in the 
Lagrangian, since the coupling of a collinear 
mode to the on-shell scalar causes the scalar propagator to 
become hard. These hard effects are already integrated out, and 
appear as effective production and decay vertices (see the following 
subsection). Finally, we must add an effective Lagrangian for 
the collinear neutrino field. Adopting a frame where the electron and 
neutrino collide head-on, the neutrino Lagrangian is obtained 
from (\ref{scetLO}) by substituting $\psi_c\to\chi_{c2}$, 
interchanging $\nm \leftrightarrow \np$, and replacing 
all covariant derivatives by ordinary derivatives. At the level 
of the leading Lagrangian the neutrino is non-interacting. 

To summarize, Feynman diagrams in resonant 
scattering involve contributions from the following three momentum 
regions:
\beq
\begin{array}{rl}
\mbox{hard (h):}\qquad & p \sim M \\
\mbox{soft (s):} \qquad& p \sim M\delta \\
\mbox{collinear (c1):} \qquad& p_\perp \sim M\delta^{1/2}, 
\, n_+ p \sim M, \, n_- p \sim M\delta
\end{array}
\eeq
(We note that what we call ``soft'' here is called ``ultrasoft'' 
in \cite{BCDF} and much of the literature on soft-collinear effective 
theory.) 
In the general case several types of collinear modes are required, one for 
each direction defined by energetic particles in the initial and 
final state. For the inclusive line shape we calculate 
the forward scattering amplitude, so no direction is distinguished in the 
final state. We then need two sets of collinear 
modes, one for the direction of the incoming electron, labeled 
by ``c1'' (or often simply ``c''), 
the other for the direction of the incoming neutrino (labeled ``c2''). 
When the hard fluctuations are integrated out, 
the effective Lagrangian 
is built from the soft heavy scalar $\phi_v$, a soft and 
collinear photon, $A_s$ and $A_{c1}$, respectively,  
a collinear and a soft electron, $\psi_{c1}$ and $\psi_s$, 
and a c2-collinear and soft neutrino, $\chi_{c2}$ and $\chi_s$. 
(In complete generality, we should also introduce the collinear 
fields $\psi_{c2}$, $A_{c2}$ and $\chi_{c1}$, but they appear 
only in highly suppressed terms, so we can ignore them here.)  
The effective interactions 
of these modes are described by the sum of ${\cal L}_{\rm HSET}$ 
and the soft-collinear effective Lagrangians for the two 
directions. In order to 
describe resonant scattering with a desired accuracy one has to match the 
Lagrangian to this accuracy, and calculate the scattering amplitude 
in the effective theory. The relevant terms can be identified a priori 
by applying the $\delta$ power counting rules. 

\subsubsection{Effective vertices}

The effective Lagrangian misses an essential piece of physics. 
Since ${\cal L}_{\rm HSET}$ does not contain collinear fields, and 
since ${\cal L}_{\rm SCET}$ does not contain the heavy scalar field, 
the two can interact only through soft modes. In particular, there 
is no vertex in the Lagrangian that allows the production of the 
scalar in the scattering of energetic particles.

Such vertices cannot be included in the effective Lagrangian for soft and 
collinear terms as  
interaction terms, because they contribute to the scattering matrix 
element in the effective theory 
only in a very specific pattern. For instance, a vertex with fields 
$\phi_v\bar\psi_{c1}\chi_{c2}$ can occur exactly twice, once
for the production of the scalar, and once for its decay. Multiple
insertions are not compatible with the kinematic restrictions on 
the process, which allow only one nearly on-shell scalar line 
in any diagram by energy conservation. 
Note that a diagram in the full theory 
may of course have many internal scalar lines. However, these will 
generally be far off-shell (``hard'') and these hard effects 
are included in the coefficient functions of the Lagrangian and 
effective vertices. What we are discussing here is the structure 
of possible diagrams/scattering processes after hard effects 
are integrated out. 

In more technical terms the interaction of two collinear modes with
opposite directions, here the electron and the neutrino, produces a
hard momentum configuration with invariant mass squared {\em of order}
$M^2$, and therefore cannot be included in the effective Lagrangian. What we
actually need is that the momenta of the oppositely moving collinear
modes are pre-arranged (by the experimenter who sets up the beam
energy to be near resonance) to produce a configuration with invariant
mass squared {\em equal} to $M^2$ within a small amount of order
$M^2\delta\sim s-M^2$. To account for these configurations we
introduce production and decay vertices (``effective vertices''). The
leading-order vertex is simply the original Yukawa coupling expressed
in terms of the fields in the effective theory,
\begin{equation}
\label{JLO}
J(x) = e^{-i\hat M v\cdot x}\,y\,
\big[\phi_v\bar \psi_{c1}\chi_{c2}\big](x).
\end{equation}
In higher order in $\alpha$ or $\delta$ a larger set of operators -- 
not necessarily local -- 
is generated by integrating out hard quantum fluctuations and the
coefficient function will receive corrections. 
(This will be seen in more detail in Section~\ref{sec:vertices}.) 
However, these operators have in common that they contain the field 
$\phi_v$ exactly once. 

The scattering may also occur without the production of the scalar near its 
mass-shell. In our toy theory this still requires an intermediate
scalar line, since the neutrino has only Yukawa interactions. 
The scalar may be off-shell, because the electron has radiated 
an energetic (hard or collinear) photon before it hits the neutrino. 
In this case the invariant mass of the colliding electron-neutrino system 
is of order $M^2$ but not near $M^2$, producing a non-resonant
scalar. In the
effective theory this process is represented by production-decay
operators, which do not contain $\phi_v$ fields. The simplest 
operator of this type has the coupling and field structure 
\begin{equation}
\label{Tstruct}
T(x) \sim |y|^2\,
\big[(\bar\psi_{c1}\chi_{c2})(\bar\chi_{c2}\psi_{c1})\big](x).
\end{equation}
In general, non-resonant scattering includes all ``background
processes'', which produce one of the final states under 
consideration.

\begin{figure}[t]
   \vspace{-5cm}
   \epsfysize=27cm
   \epsfxsize=20cm
   \centerline{\epsffile{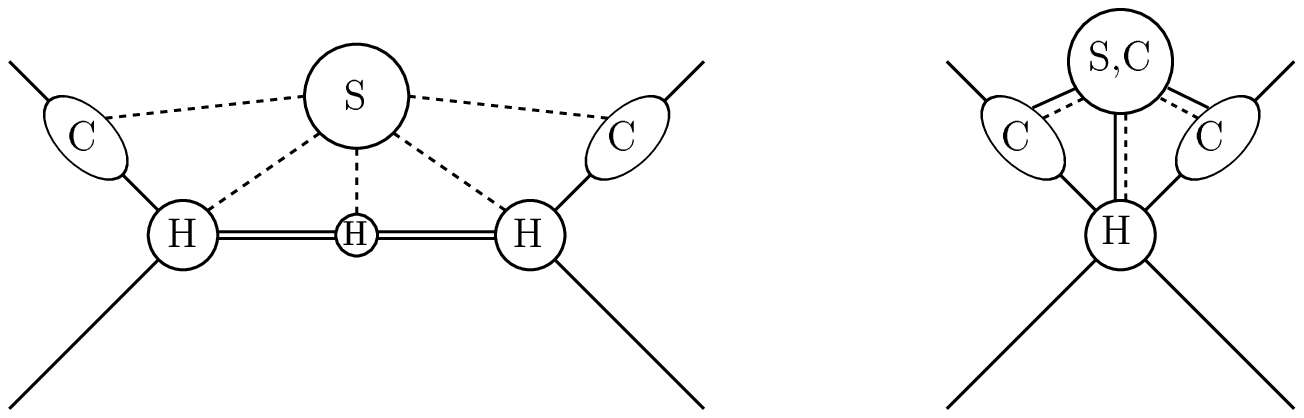}}
   \vspace*{-17.3cm}
\centerline{\parbox{14cm}{\caption{\label{fig:skeleton}\small  
Reduced diagram topologies in $2\to2$ scattering 
near resonance. Left: resonant scattering. Right: non-resonant  
scattering. See text for explanations.}}}
\end{figure}

With these preparations we obtain the following formula for the 
forward $2\to2$ scattering amplitude in the effective theory:
\begin{equation}
\label{master}
i\,{\cal T} = \sum_{m,n} \,\langle \bar\nu e|\int d^4 x\,
T\big\{i J_m^\dagger(0) i J_n(x)\big\} |\bar\nu e\rangle + 
\sum_k \,\langle \bar\nu e|i T_k(0) |\bar\nu e\rangle,
\end{equation} 
where in the first term ``$T$'' denotes time-ordering, and 
where the sums extend over the sets of effective vertices of the two 
types, $J$ or $T$. The matrix elements on the right-hand side 
are evaluated with the effective Lagrangian for soft and collinear 
fields. This result is depicted in graphical form in 
Figure~\ref{fig:skeleton}. The ``resonant'' scattering term on 
the left-hand side involves a production and decay vertex, and collinear 
interactions at the electron legs, which connect to the production and
decay vertex. Soft (and only these) fields  
can connect to all parts of the diagram. The double line denotes
the heavy resonance, solid lines collinear modes and dashed lines 
soft modes. A single solid (dashed) line may represent a collection
of several collinear (soft) fields, depending 
on the structure of the effective interaction. Collinear fields cannot
be exchanged from the left to the right of a resonant diagram, 
since this would cause the scalar to be off-shell. The corresponding 
configurations are included 
in the ``non-resonant'' scattering topology on the right-hand side of 
the figure. This topology 
does not involve a resonant heavy scalar, and both
soft and collinear fields can be exchanged across the diagram.

The various terms in (\ref{master}) can be ordered according to 
their scaling with $\delta$ and $\alpha$, thus allowing for a
determination of the terms needed for a calculation to some accuracy 
before performing any computation.  We will now estimate the
importance of the resonant and non-resonant contributions by a 
simple power counting argument, taking as representatives the leading 
interactions of each type. Combining the scaling of the fields 
we find $J\sim y\,\delta^{5/2}$ from (\ref{JLO}). Furthermore 
$x$ is soft, since $x$ represents the distance of 
order $1/\delta$ over which the resonant scalar propagates, 
hence $d^4 x$ in (\ref{master}) counts $1/\delta^4$. The
normalization of states implies that an external collinear particle 
counts as $1/\delta^{1/2}$, so $|\bar\nu e\rangle\sim 1/\delta$. 
Putting everything together, we obtain 
\begin{equation}
\label{count1}
\sigma \sim \frac{1}{s}\,\mbox{Im}\,{\cal T} \sim \frac{|y|^2}{\delta} 
\sim \frac{|y|^2}{M\Gamma},
\end{equation}
which is the expected result. The leading non-resonant 
scattering operator $T$ in (\ref{Tstruct}) scales as $|y|^2\,\delta^2$. 
The matrix element $\langle \bar\nu e|i T(0) |\bar\nu e\rangle$ 
is non-vanishing only at order $g^2$, since the overlap of $T$ with
the external states requires the emission of an energetic photon. 
Adding the counting of the states, we find that 
the contribution to ${\cal T}$ is of order $g^2|y^2|$. This is
suppressed by a factor $g^2\Gamma/M$ relative to (\ref{count1}), 
hence establishing that non-resonant topologies contribute only at 
next-to-next-to-leading order in the combined expansion in 
$\alpha $ and $\delta$. We shall discuss these effects in 
some detail in Section~\ref{sec:pd4}.

In this discussion of the non-resonant contribution we assumed that 
the $(\bar\psi_{c1}\chi_{c2})$ operator in $T$ describes an
electron-neutrino configuration with invariant mass far from $M^2$. We
shall see below that the applications of the equation of motion to
bring the scalar effective Lagrangian into its canonical form generate
a local four-fermion operator identical to $T$, but with 
$(\bar\psi_{c1}\chi_{c2})$ describing an
electron-neutrino configuration with invariant mass near $M^2$. 
This arises because the resonant propagator is canceled by the
rearrangements implied by the equation of motion. In this case $T$ has
a non-vanishing tree-level matrix element and contributes already at 
next-to-leading order. 

\subsubsection{Soft and collinear fluctuations of collinear modes}

The complication mentioned in the previous paragraph, as well as the
complication of excluding ``by hand'' collinear loop contributions to
the matrix element of $T\left\{J^\dagger(0)J(x)\right\}$, which 
connect the initial and final state, can be avoided, if we distinguish
modes that differ from the external collinear momenta only by a soft
momentum from those which differ by a collinear momentum. Both modes
have virtuality $M\delta^{1/2}$ and have been called ``collinear'' up
to now. However, the former can be assigned momentum 
$\hat M n_-/2+k$, where $k$ is soft and the large collinear 
component $\hat M n_-/2$ is {\em fixed}. Only the latter represent
collinear fluctuations associated with collinear loop momenta. 

Just as for the heavy scalar with momentum close to $\hat M v$, 
it is useful to extract the fixed large momentum from the 
field, and to define 
\begin{equation}
\psi_{n_-}(x) = e^{i\hat M/2\,(\nm x)}\,{\cal P}_+\psi_{c1}(x)
\end{equation}
for soft fluctuations around the external momentum. (${\cal P}_+$ 
projects on the positive frequency part of $\psi_{c1}$.) 
The new field $\psi_{n_-}(x)$ has only variations over distances 
of order $1/\delta$. For the collinear fluctuations we keep the 
field $\psi_{c1}(x)$. Similarly we define the field 
$\chi_{n_+}(x)$. 

The collinear effective Lagrangian now assumes a more complicated
expression, since we distinguish the interactions of the two types of
collinear fields. However, it is now possible to add the effective
vertices as interaction terms in the Lagrangian, since the kinematic
distinction of configurations with virtualities of order and near $M^2$
is implemented at the level of the fields. In particular, the
effective vertex operator $J(x)$ is now given by
\begin{equation}
y\,[\phi_v\bar\psi_{n_-}\chi_{n_+}](x),
\end{equation}
but there is no interaction vertex of this form with $\bar \psi_{n_-}$
replaced by $\bar\psi_{c1}$. For the four-fermion operators,
we must distinguish 
$(\bar\psi_{c1}\chi_{n_+})(\bar\chi_{n_+}\psi_{c1})$
from $(\bar\psi_{n_-}\chi_{n_+})(\bar\chi_{n_+}\psi_{n_-})$.
Since the external collinear state can be created or
destroyed only by the softly fluctuating 
collinear fields and their conjugates, the 
latter operator has a tree-level matrix element, but the former does
not, because it requires interaction vertices of the type 
$\bar \psi_{n_-}\psi_{c1} A_{c1}$ from the collinear Lagrangian to
generate a non-vanishing overlap with the external state.

\subsubsection{Integrating out collinear fluctuations}

The formalism developed up to this point contains propagating soft 
and collinear modes in the effective theory. We can go one step
further and integrate out collinear {\em fluctuations}, that is 
we integrate out loops with momenta that scale as collinear 
momenta. This can be done perturbatively as long as the couplings at 
the scale $M\delta^{1/2}$ are small. We now sketch the structure of 
the resultant effective theory. At the level of the calculations 
in the remainder of the paper there is in fact no difference between 
the two descriptions before and after integrating out collinear loops,
since the collinear loop integrals we encounter all vanish in
dimensional regularization.

Technically we introduce two distinct collinear fields as described
above and eliminate the fields that describe collinear fluctuations
from the effective Lagrangian. 
With only soft modes remaining in the effective theory, the 
effective Lagrangian is constructed from the 
soft electron and photon field, and the fields $\phi_v$, $\psi_{n_-}$,
and $\chi_{n_+}$, all scaling as $\delta^{3/2}$. 
The scattering amplitude can be computed from 
\begin{equation}
{\cal L}={\cal L}_{\rm HSET}+{\cal L}_- + {\cal L}_+ + 
  {\cal L}_{\rm int},
\end{equation}
where ${\cal L}_{\rm HSET}$ describes the soft interactions of the 
heavy scalar as before, and ${\cal L}_-$, ${\cal L}_+$ are bilinear 
in the fields $\psi_{n_-}(x)$ and $\chi_{n_+}(x)$, respectively. The 
soft-collinear Lagrangian ${\cal L}_{\rm SCET}$ is non-local due to the  
presence of $1/(i\np \partial)$, but these non-localities disappear, when 
collinear fluctuations are integrated out, and the large component 
of the energetic fields is ``frozen'' to $\hat{M}/2$ (modulo soft 
variations). For instance, consider the leading-power SCET 
Lagrangian (\ref{scetLO}) for the electron field, 
which matches at tree level to the expression 
\begin{equation}
\bar{\psi}_{n_-} \!\left(i n_- D_s +  \frac{[i \partial_\perp]^2}
{\hat M}\right) \frac{\slash{n}_+}{2} \, \psi_{n_-}
\end{equation}
in the new effective theory. 
The term with the transverse derivative squared is already suppressed 
by one factor of $\delta$ compared to the leading term, since 
derivatives on $\psi_{n_-}(x)$ count as $\delta$. 
When collinear quantum fluctuations 
are included, the set of allowed operators consists of operators 
non-local in $1/(i\nm \partial)$, since the $n_-$ component of 
the external soft momenta of a collinear loop are of the same order 
as the $n_-$ component of the collinear momentum. These new
non-localities, appearing at higher orders in $\alpha$,   
correspond to the convolution of a ``jet factor'' with the remaining soft 
matrix element. Similar considerations apply to the neutrino field. 
The fields for the soft fluctuations of the 
energetic particles now behave in many ways
similar to the field for a static heavy particle. In both cases 
the virtuality is $\hat M^2\delta$, and the interaction with soft 
modes (of virtuality $\hat M^2\delta^2$) kicks the particle off its 
classical trajectory only by a negligible amount; hence the fixed vector 
that labels the field. (The discussion here is 
reminiscent of ``large-energy effective theory'' \cite{Dugan:1990de},
except that we have in mind a situation where hard {\em and collinear}
modes are eliminated.)

As mentioned above, the production and decay 
vertices can be implemented as interaction terms ${\cal L}_{\rm int}$ 
in the Lagrangian, because the kinematic 
restriction that prevented us from adding generic interactions of 
collinear modes with different directions to the Lagrangian is now implemented 
at the level of the fields: by construction the scattering of 
the electron described by $\psi_{n_-}(x)$ and the neutrino 
described by $\chi_{n_+}(x)$ produces a mode with momentum 
$\hat M v+k$ with $k$ soft. 
The most general terms allowed for a 
$2\to2$ scattering process are of the form 
$\phi_v\bar\psi_{n_-}\chi_{n_+}$ (or its hermitian conjugate) or 
$(\bar\psi_{n_-}\chi_{n_+})\,(\bar\chi_{n_+}\psi_{n_-})$ plus 
additional soft fields. Since none of the 
fields $\phi_v$, $\psi_{n_-}$,
and $\chi_{n_+}$ is created or destroyed by the other terms in 
the Lagrangian, the terms in ${\cal L}_{\rm int}$ can only 
contribute in a topology that corresponds to a classical scattering 
process of the energetic particles and the heavy scalar. (In 
particular closed loops of only $\phi_v$, $\psi_{n_-}$ and $\chi_{n_+}$ 
lines vanish due to the structure of the poles of the propagators.) 
These are the two topologies shown in Figure~\ref{fig:skeleton} 
but with the collinear subgraphs eliminated and interpreted as 
parts of the production and decay vertices. The power counting goes as
follows: since $\int d^4 x \phi_v\bar\psi_{n_-}\chi_{n_+}\sim
\delta^{1/2}$, each insertion of this interaction vertex 
gives a factor of $\delta^{1/2}$. (Even though the effective
theory contains only a single scale, we must count powers of $\delta$
rather than dimensions due to our unconventional $2\hat M$
normalization of the non-relativistic scalar kinetic term.) 
The insertion of a four-fermion operator carries a
$\delta^2$ suppression factor, which results in a $\delta$ suppression
relative to the resonant contribution, not counting factors 
of coupling constants.

\subsection{Leading-order line shape}
\label{sec:loLineShape}

We now perform the (trivial) leading-order calculation of 
the inclusive line shape of the resonance. From the foregoing it
follows that we need only the leading-order Lagrangian 
(\ref{heavyLO}) together with the leading vertex $J$ in 
(\ref{JLO}). The calculation consists of two steps: matching 
the effective Lagrangian to the ``full'' theory to leading 
order, and the calculation of the forward scattering amplitude in 
the effective theory. At the end, we use the optical theorem to obtain
the line shape. 

\subsubsection{Matching} 

Since $i v\partial$ is of order $\delta$ on soft fields, 
we see from (\ref{heavyLO}) 
that we need $\Delta$ to order $\delta\sim\alpha$. This implies 
that $\Delta$ must be chosen so as to reproduce the (renormalized) 
full theory propagator $i/(p^2-\hat M^2-\Pi(p^2))$ near the resonance 
pole with this accuracy. We discuss the matching condition in more 
technical terms in Section~\ref{sec:heavy}. At leading order, one finds 
$\Delta^{(1)} = \Pi^{(1)}(\hat M^2)/\hat M$,
where $\Pi^{(1)}(\hat M^2)$ is the renormalized 
one-loop self-energy evaluated at $p^2=\hat M^2$. 

All calculations will be done with dimensional regularization with 
$d=4-2\epsilon$, where 
the loop integration measure is defined as 
\begin{equation}
[dl] = \left(\frac{\mu^2 e^{\gamma_E}}{4\pi}\right)^{\!\epsilon}\,
\frac{d^dl}{(2\pi)^d}.
\end{equation}
With this convention 
the $\overline{\rm MS}$ scheme corresponds to subtracting all poles 
in $\epsilon$ minimally. In our toy model a straightforward 
computation of four diagrams
\vskip0.2cm
\begin{equation}
\label{PILO}
  -i \,\Pi^{(1)} = 
  \unitlength .7pt\SetScale{0.7}
  \begin{picture}(100,30)(0,30)
    \DashLine(0,30)(50,30){4}
    \PhotonArc(50,45)(15,0,359){2}{8}
    \DashLine(50,30)(100,30){4}
  \end{picture}
+
  \unitlength .7pt\SetScale{0.7}
  \begin{picture}(100,30)(0,30)
    \DashLine(0,30)(25,30){4}
    \ArrowArc(50,30)(25,180,0)
    \ArrowArc(50,30)(25,0,180)
    \DashLine(75,30)(100,30){4}
  \end{picture}
+
  \unitlength .7pt\SetScale{0.7}
  \begin{picture}(100,30)(0,30)
    \DashLine(0,30)(25,30){4}
    \PhotonArc(50,30)(25,0,180){2}{6}
    \DashCArc(50,30)(25,180,0){4}
    \DashLine(75,30)(100,30){4}
  \end{picture}
  \vspace{30 \unitlength}
+
  \unitlength .7pt\SetScale{0.7}
  \begin{picture}(100,30)(0,30)
    \DashLine(0,30)(50,30){4}
    \Line(45,35)(55,25)
    \Line(45,25)(55,35)
    \DashLine(50,30)(100,30){4}
  \end{picture}
\end{equation}
gives 
\begin{equation}
\frac{\Delta_{\overline{\rm MS}}^{(1)}}{\hat M} = 
\frac{\alpha_y}{4\pi}\left(2\ln\frac{\hat M^2}{\mu^2}-4-
  2i \pi \right) +\frac{\alpha_g}{4\pi}\left(-3\ln\frac{\hat M^2}{\mu^2}+
  7\right)
\end{equation}
in the $\overline{\rm MS}$ scheme. (Note that the one-loop 
diagram proportional to the scalar self-coupling $\lambda$ counts as 
$\alpha^2$ and therefore does not contribute to $\Pi^{(1)}$.) 

The renormalization scheme dependence of $\Delta$ is related to the 
mass renormalization convention in the underlying theory, if we 
choose the parameter $\hat M$ in the effective theory to be the
renormalized mass of the full theory. The effective theory formalism 
does not rely on a specific renormalization convention. We will 
use the $\overline{\rm MS}$ convention and the pole
mass convention to illustrate this point. The pole mass $M$ is defined 
through the location of the singularity,
\begin{equation}
\bar s = M^2-i M\Gamma
\end{equation}
in the scalar propagator. Writing 
$M=\hat M+\delta \hat M$, and using $\Delta^{(1)} = \Pi^{(1)}(\hat
M^2)/\hat M$, the relations 
\begin{equation}
\label{1looprel}
\delta \hat M^{(1)} = \frac{1}{2}\,\mbox{Re} 
\, \Delta^{(1)},\qquad
\Gamma^{(1)} = -\mbox{Im}\, \Delta^{(1)}
\end{equation}
hold at the one-loop order. Note that $\hat M=\hat M(\mu)$, but we do 
not indicate the scale-dependence explicitly.

\subsubsection{Forward scattering amplitude}

The forward scattering amplitude is given at leading order in the 
effective theory (not distinguishing explicitly between collinear and 
soft fluctuations of the collinear fields) by
\begin{eqnarray}
\label{TLO}
i\,{\cal T}^{(0)} &=& \int d^4 x\,e^{-i \hat M v\cdot x}\,\langle \bar\nu e|
T\big\{i y^* \,\big[\phi_v^\dagger \bar \chi_{c2}\psi_{c1}\big](0)\, 
i y\big[\phi_v\bar \psi_{c1}\chi_{c2}\big](x)\big\}|\bar\nu e
\rangle_{\rm LO}
\nonumber\\[0.3cm]
&=&   
  \unitlength .7pt\SetScale{0.7}
  \begin{picture}(100,30)(0,30)
    \ArrowLine(30,30)(0,0)
    \ArrowLine(0,60)(30,30)
    \GCirc(30,30){2}{0}
    \Line(30,31)(70,31)
    \Line(30,29)(70,29)
    \GCirc(70,30){2}{0} 
    \ArrowLine(100,0)(70,30)
    \ArrowLine(70,30)(100,60)
  \end{picture}
\vspace{3mm}
= i^2 y y^* \, [\bar{u}(p)v(q)]\,\frac{i}{2\hat M(v \cdot k
-\Delta^{(1)}/2)}\,[\bar{v}(q)u(p)], 
\\\nonumber
\end{eqnarray}
where the double line denotes the resonant propagator defined 
by the effective Lagrangian ${\cal L}_{\rm HSET}$, and 
$k=p+q-\hat M v$. We take $v=(1,\vec{0})$ and the external momenta 
$p=\sqrt{s}/2\,(1,0,0,-1)=\sqrt{s}/2\,\nm$, 
$q=\sqrt{s}/2\,(1,0,0,1)=\sqrt{s}/2\,\np$, so $v\cdot k =
\sqrt{s}-\hat M$. Note that by construction 
the effective propagator includes 
the geometric series of all one-loop self-energy insertions, evaluated
at threshold according to the definition of $\Delta^{(1)}$. 
Performing the polarization average the forward scattering amplitude 
reads 
\begin{equation}
\label{LOT0}
{\cal T}^{(0)} = -\frac{y y^*\,s}{4\hat M (\sqrt{s}-\hat M-\Delta^{(1)}/2)}.
\end{equation}
Note that because we used spinors $u(p)$ etc. for the external 
states, ${\cal T}^{(0)}$ contains next-to-leading order terms from 
the factor $s=\hat M^2+[s-\hat M^2]$ in the numerator. For a strict 
expansion it would be more appropriate to use the spinors 
$u(\hat M\nm/2)$ etc., in which case 
${\cal T}^{(0)}$ would be strictly leading order, but in practice it
may be convenient not to perform this trivial expansion in order to
reduce the number of terms. 

\subsubsection{Inclusive line shape}

The inclusive line shape is related to the forward scattering
amplitude by the optical theorem, which gives
\begin{eqnarray}
\label{lineLO}
\sigma^{(0)} &=& \frac{1}{s}\,\mbox{Im}\,{\cal T}^{(0)} = 
\frac{y y^*}{4\hat M} \,\frac{\Gamma^{(1)}/2}
{(\sqrt{s}-[\hat M+\delta\hat M^{(1)}])^2+{\Gamma^{(1)}}^2/4}
\nonumber\\
&=& \frac{\pi\alpha_y^2}{4} \,\frac{1}
{(\sqrt{s}-[\hat M+\delta\hat M^{(1)}])^2+(\alpha_y\hat M/4)^2},
\end{eqnarray}
where we used (\ref{1looprel}).
The line shape has the form of a 
Breit-Wigner distribution in the center-of-mass energy $\sqrt{s}$. 
This is in fact the universal line shape that appears in the 
leading-order approximation to any resonant $\bar\nu e$ scattering 
process in our toy model, 
but the line shapes will depend on the final state beyond 
this approximation.  (It is a matter of convention whether
we divide by $s$ to obtain the cross section from the forward
scattering amplitude, or expand this factor
around $\hat M^2$. Here we keep $1/s$ unexpanded, so that the factor
of $s$ in the numerator of (\ref{LOT0}) is canceled, and the
leading-order line shape assumes an exact Breit-Wigner distribution.) 

In the presence of fields for unstable particles unitarity and 
the optical theorem apply to the scattering matrix defined on 
the Hilbert space of stable particle in- and out-states only, since 
the resonance does not correspond to an asymptotic particle 
state \cite{Veltman:th}. The optical theorem can be interpreted 
as taking the sum over all ``cuts'' of a diagram, where now 
the line for the unstable particle propagator must not be cut. 
For the leading order diagram in (\ref{TLO}) this means that 
cutting the effective propagator represents the sum over all 
possible cuts of the one-loop self-energy insertions implicitly 
contained in the double-line propagator. This corresponds to 
cutting the second diagram in (\ref{PILO}), which is responsible for
the leading-order decay width $\Gamma^{(1)}$ of the scalar. 

We remark that the effective Lagrangian is not real, since the hard 
coefficient functions have imaginary parts, but this does not lead to a 
non-unitary time evolution. Consider, for instance, the diagram in 
(\ref{TLO}). Unitarity requires an amplitude that corresponds to the 
``square'' of this diagram, but the effective theory does not 
contain diagrams with closed electron-neutrino loops. Rather the 
corresponding (short-distance) effect is included in the complex 
coefficient function. Matching the effective theory to the underlying 
theory automatically guarantees that the combination of effective 
theory diagrams and complex couplings reproduces the unitary 
time-evolution to the desired accuracy. 

\begin{figure}[t]
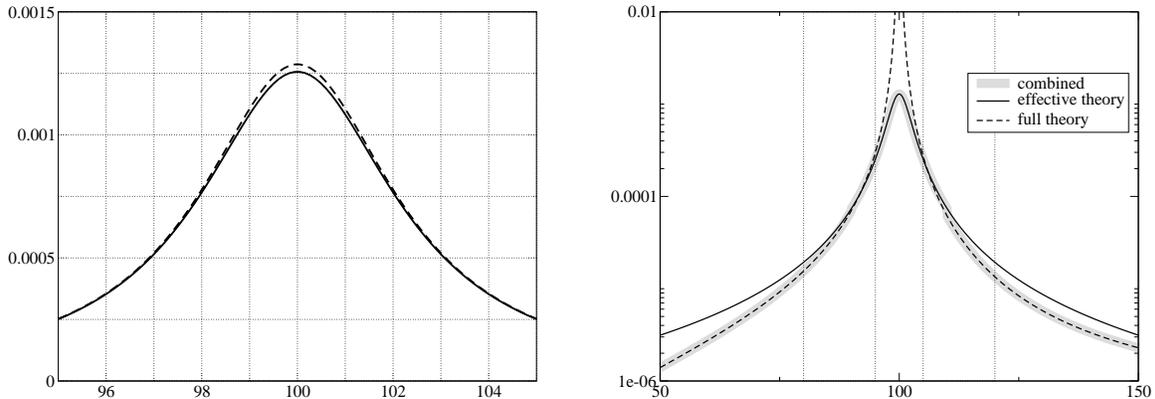

  \vspace*{-1.2cm}
  \unitlength 1cm
  \begin{center}
  \begin{picture}(13.4,7)
  \put(-1.2,0){\includegraphics{born.msbar-onshell.eps}}
  \put(6.8,0){\includegraphics{born.hard-eff.eps}}
  \end{picture}
  \end{center}
  \centerline{\parbox{14cm}{\caption{\label{fig:line-shape}\small  
  Left panel: line shape (in $\mbox{GeV}^{-2}$) in the pole (solid) and
  $\overline{\rm MS}$ scheme (dashed) as a function of the
  center-of-mass energy (in GeV). 
  Right panel: Leading-order line shape  (in $\mbox{GeV}^{-2}$) 
  as a function of
  center-of-mass energy (in GeV) in the effective theory
  (solid) and the cross section off resonance in the full theory
  (dashed). The thick grey curve shows the leading-order line-shape 
  with the two curves matched appropriately.}}}
\end{figure}

The calculation shows that if the result is represented in two
different mass renormalization schemes, the masses must be related to
one-loop accuracy for a consistent change of conventions. For
instance, if the pole mass $M=100\,$GeV is known from elsewhere, 
the $\overline{\rm MS}$ mass for $\mu=\hat M$ 
is $\hat M=M-\delta\hat M^{(1)}=
98.8\,$GeV, where we use $\alpha_y=\alpha_g=0.1$ to illustrate 
the numerics. From (\ref{lineLO}) we see that the denominator 
contains $\sqrt{s}-M$ in {\em any} scheme to one-loop accuracy, 
but there is a residual scheme dependence (and scale dependence, 
since $\hat M$ depends on $\mu$) due to the width 
$\alpha_y\hat M/2$. This is a NLO effect, which will be 
reduced by a higher-order calculation, as will be the dependence on 
the renormalization convention of the coupling constants. 
The line shape in the pole and $\overline{\rm MS}$ scheme is shown 
in the left panel of Figure~\ref{fig:line-shape}. 

The effective theory calculation (\ref{lineLO}) is valid only 
near resonance when $\sqrt{s}-\hat M$ is small, since the 
expansion inherent to the construction neglects corrections of 
this order. In other words, near resonance the corrections are 
of order $\alpha$ or $\delta$, but the second quantity becomes 
large far off the resonance. To obtain an adequate description of 
the line shape in the entire kinematical range, the effective theory
calculation must be matched to a full theory
calculation off resonance. The point to note here is that the 
full theory calculation of the forward scattering amplitude 
can be done with ordinary perturbation theory, since no kinematic 
enhancements invalidate it off resonance. In particular, the 
optical theorem is applied to intermediate states including 
the scalar particle as if it were stable. 

The off-resonance cross section is only of order $\alpha^2$ compared
to order one  near resonance. The result takes the form 
\begin{equation}
\label{fullLO}
\sigma= \frac{\pi\alpha_y^2 \,s}{(s-\hat M^2)^2}
\left(f_y(\hat M^2/s)+\frac{\alpha_g}{\alpha_y}\,\theta(s-\hat M^2) \,
f_g(\hat M^2/s) \right)
\end{equation}
with 
\begin{eqnarray}
f_y(z) &=& 1,
\nonumber\\
f_g(z) &=& (1-z)\left\{(1+z^2)\,\ln\frac{s (1-z)^2}{\nu^2}-2 z \right\}.
\end{eqnarray} 
The ``partonic'' cross
section has a collinear divergence, which has been subtracted
according to the $\overline{\rm MS}$ prescription. The physical cross
section follows from convoluting the above result with an electron
distribution function (in the electron) that depends on the
subtraction scale $\nu$. The argument of the logarithm suggests that
the collinear factorization scale should be taken to be of order
$(s-\hat M^2)/\hat M\sim \hat M\delta$. In the right panel of
Figure~\ref{fig:line-shape} we show the leading-order computation in
the effective theory, and in the full theory off resonance, evaluated
at the running scale $\nu=\sqrt{s} \,(1-z)$. The
effective theory calculation correctly represents the line shape near
the resonance with relative accuracy $\alpha$. The peak height is of
order $1$. As the calculation is extrapolated off resonance the cross
section decreases and becomes of order $\alpha^2$. However, in this
region the relative error on the effective theory calculation is of
order unity. The full theory calculation correctly represents the
off-resonance cross section of order $\alpha^2$ with relative accuracy
$\alpha$. As the calculation is extrapolated towards the resonance,
the relative error becomes of order unity. (In fact, the cross section
diverges at $s=\hat M^2$ and the error becomes arbitrarily large.) The
two calculations must be matched in the intermediate region where
$\delta \sim (\sqrt{s}-\hat M)/\hat M$ is small enough for both
calculations to be valid. The existence of this intermediate region is
indicated in the figure by the vertical lines to the left and the 
right of the resonance peak.  We also verify from 
(\ref{lineLO}) and (\ref{fullLO})
that the two calculations agree analytically in this intermediate
region. (For completeness we mention that the partonic cross section 
proportional to $\alpha_y\,\delta(s-\hat M^2)$ should be included in
the convolution with the electron distribution function, which ensures
that the off-resonance cross section is $\nu$-independent, and sums 
large logarithms of $\sqrt{s}$ over the electron mass. Since this
procedure is standard, we presented the partonic cross section at the
running scale, at which no large logarithms occur.)

\section{The line shape at next-to-leading order \label{sec:nlo}}

Following the formalism set up in the previous section we now discuss
the matching calculations in more detail. We construct the effective
Lagrangian by matching on-shell Green functions and present the
explicit form of all terms of the effective Lagrangian that are needed
for the evaluation of the line shape at next-to-leading order.  
At the end of this section we also compute 
the scattering matrix element in the effective theory to 
next-to-leading order and obtain the complete NLO line shape. 

The matching procedure involves three steps. Compute the 
renormalized on-shell Green functions in the full theory up to the
required order in $\alpha$. Evaluate the same
quantity in the effective theory. Determine the hard matching 
coefficient so that the two calculations agree within a specified 
accuracy. The Green functions in the full theory are calculated using
conventional weak-coupling perturbation theory. In general, 
we need an infrared regularization, and the same regularization 
must be used when computing the corresponding quantity in
the effective theory. The result for the matching coefficient 
is independent of the IR regularization. The calculations become 
particularly simple with dimensional regularization, 
since the effective theory loop diagrams 
vanish in this scheme. This is simply a consequence of
the fact that the matching is done on-shell, which renders 
the corresponding integrals scaleless. However, the matching
coefficients depend on a factorization scheme, since the Green functions in 
the effective theory also exhibit ultraviolet singularities. 
Thus they have to be UV-subtracted and
the choice of the renormalization scheme for the effective theory
amounts to the choice of the factorization scheme.

In practice, we simplify the matching procedure by computing directly
the hard part of the renormalized on-shell Green functions in the full
theory, using the method of regions~\cite{Beneke:1998zp}. This also
applies to the $Z$-factors that multiply the amputated on-shell Green
function. Even though the method of
regions forces us to use dimensional regularization to define the 
factorization scheme, we are still free
to use another infrared regulator, such as a photon mass, 
in which case the loop diagrams in the effective theory are 
generally non-vanishing. 
However, the result of the hard part is independent of the
infrared regularization of the full theory, but it contains infrared
$1/\eps$ poles due to the separation into hard and soft
contributions. These singularities coincide with the above-mentioned UV
singularities in the effective theory. We choose to subtract them
minimally and thereby define the factorization scheme and the
renormalization scheme for the effective theory.

\subsection{\boldmath Matching ${\cal L}_{\rm HSET}$}
\label{sec:heavy}

For the calculation of the line shape at leading order it is
sufficient to consider the first term of ${\cal L}_{\rm HSET}$ 
[Eq.~(\ref{heavyLO})], with the matching coefficient $\Delta$ known to order
$\alpha\sim \delta$. At next-to-leading order we will need all terms of
order $\{\alpha^2, \alpha\delta,\delta^2\}\times  
\phi^\dagger_v \phi_v$. Furthermore, at NLO there will also
be soft photon loop diagrams in the effective theory, so we will need to
include the corresponding terms in the Lagrangian. 
Since these diagrams will be 
suppressed by a factor $\alpha$
with respect to the leading-order diagram, the soft photon 
vertices are only needed at leading order.

\subsubsection{The bilinear terms for the unstable field}
\label{sec:bilinear}

The bilinear terms of the effective Lagrangian are determined 
by the dispersion relation for the free-particle states. The field 
$\phi_v$ describes a heavy particle near mass-shell, so we need 
the equivalent of the heavy-quark effective Lagrangian for a scalar
field. The only complication is that the scalar is unstable, so the 
notion of a free-particle state is not really defined. In this case 
the bilinear terms are constructed so as to reproduce the 
two-point function near the resonance pole. 

In the underlying theory the full renormalized propagator for the
unstable particle is given by $i (s-\hM^2-\Pi(s))^{-1}$, where
$-i\,\Pi(s)$ corresponds to the amputated 1PI graphs, including
counterterms. Recalling $P= \hM v +k$ the
propagator near the resonance pole can be written as
\be
\frac{i\, R_\phi}{P^2 - \bar s} = 
\frac{i\, R_\phi}{2 \hM v k + k^2 -(\bar s -\hM^2)}, 
\label{eq:propagator}
\ee
where we denoted the complex pole of the propagator by $\bar s$ and
the residue at the pole by $R_\phi$. Defining 
\be
\Delta \equiv \frac{\bar s -\hM^2}{\hM},
\label{eq:Deltadef}
\ee
we solve $P^2=\bar s$ in the form
\be
\label{eq:eqofmotion}
vk = -\hM + \sqrt{\hM^2 + \hM \Delta -k_{\top}^2}
= \frac{\Delta}{2}-\frac{\Delta^2+4 k_\top^2}{8\hM} + 
{\cal O}(\delta^3),
\ee
where for any vector $a_\top^\mu\equiv a^\mu - (v a)\, v^\mu$. 
The second solution has $vk \sim \hat M$ and is irrelevant. 
From this we obtain the effective Lagrangian
\be
\label{eq:Leff_phiphi3}
  \cL_{\phi\phi} =  
 2 \hM  \phi_v^\dagger\, 
        \left( i v \cdot D_s - \frac{\Delta}{2} \right) \phi_v  +  
 2 \hM  \phi_v^\dagger\,
        \left( \frac{(i D_{s\top})^2}{2\hM} +
               \frac{\Delta^2}{8 \hM} \right) \phi_v 
 + \ldots\>.
\ee
The terms in the second bracket are suppressed by one factor of
$\delta$ relative to the leading terms. Several comments are in 
order:

(i) The field $\phi_v$ is a pure destruction field. Even if the 
original scalar field had been neutral (real), the corresponding 
effective scalar field would have been complex, 
because it contains only the 
destruction part of the relativistic field. We keep a factor 
$2\hM$ in the kinetic terms to preserve the canonical mass dimension 
1 for a scalar field. 

(ii) Since the full Lagrangian is gauge invariant and so is 
the separation into hard and soft contributions, the effective  
Lagrangian is also gauge invariant. Therefore, the
interactions of the unstable field with soft photons can be taken
into account by replacing ordinary derivatives by covariant
derivatives. This has been done in (\ref{eq:Leff_phiphi3}). 

(iii) The quantity $\Delta$ plays the role of a matching coefficient. 
For a stable particle $\Delta$ vanishes in the pole mass scheme and 
is referred to as ``residual mass'' in a general scheme. In the
unstable case $\Delta$ is complex and always non-vanishing. It is 
$\Delta$ that prevents the propagators of the effective theory
diagrams from becoming singular. For the interpretation of $\Delta$ 
as a matching coefficient to be consistent,  
$\Delta$ must be given entirely by hard fluctuations. This is 
related to the fact that the location of the pole of the 
propagator (\ref{eq:propagator}) is infrared-finite. Alternatively, 
we may note that quantum corrections to the effective theory 
propagator do not modify the location of the pole to any order in 
perturbation theory, so the full self-mass must already be contained 
in the coefficient function $\Delta$. Further, we note 
that $\Delta$ is gauge-independent, because $\bar s$ and $\hM$ are 
gauge-independent.

(iv) Computing the propagator to all orders in $\delta$ with the
effective Lagrangian  (\ref{eq:Leff_phiphi3}) does not reproduce
(\ref{eq:propagator}). Instead we obtain
\be
\label{eq:effprop}
\frac{i R_{\rm eff \phi}}{2\hM\,\left((vP)-
        \sqrt{\hM^2 + \hM \Delta -k_{\top}^2}\right)} \simeq 
\frac{i\varpi^{-1} R_{\rm eff \phi} }{P^2-\bar s}
\ee
near the resonance pole, 
where we used that the coefficients of the higher-order kinetic 
terms in the effective Lagrangian renormalize in the same way as 
the leading term due to Lorentz invariance. Besides the 
different standard residue factors, there also appears the 
factor
\be
\label{eq:Phinorm}
\varpi^{-1} \equiv \left(1+\frac{\hM\Delta-k_\top^2}{\hM^2}
\right)^{\!1/2} = 
1+\cO{\alpha,\delta},
\ee
which differs from 1 even at tree level ($\Delta=0$). 

This is due to the fact that (up to the factor $2\hM$) we use the 
standard form of a non-relativistic Lagrangian, which does not 
reproduce the normalization of a relativistic field. This is not a
problem, because the different normalization is taken into account 
in matching calculations. Whenever we compute an
amputated Green function in the effective theory we multiply every
external $\phi_v$-line by the additional wave-function normalization factor
$\varpi^{-1/2}$. This will be important in Section~\ref{sec:vertices},
where we discuss the matching of the production (decay) vertex.

\subsubsection{Two-loop computation of $\Delta$}

The matching coefficient $\Delta$ is of order $\hM \delta$ with $\delta$
defined in \eqref{eq:deltadef} and will be computed from the
perturbative expansion of the hard part of the self-energy only. 
We write the expansion in the form 
\be 
\Pi_h(s) = \hM^2 \sum_{k,l}  \delta^l \, \Pi^{(k,l)},
\label{eq:Pihard}
\ee
where it is understood that $\Pi^{(k,l)}\sim \alpha^k$. 
Since the unstable field couples to massless particles the full
self-energy is not analytic at $\hM^2$. In particular, for our model
${\rm d}\Pi(s)/{\rm d}s|_{s=\hM^2}$ has an infrared singularity. 
However, only the hard part of the self-energy enters
(\ref{eq:Pihard}). This part is constructed from the Taylor expansion 
of the Feynman integrand in $\delta$ and hence analytic at $\hM^2$, 
so the Taylor expansion (\ref{eq:Pihard}) is also well-defined.

The position
of the pole of the propagator, $\bar s$, and the hard part of the
residue at the pole, $R_{h\phi}$, can be expressed in terms of the
$\Pi^{(k,l)}$. Up to the third order we find
\beq
\frac{\bar s}{\hM^2} &=& 1+\Pi^{(1,0)}+\Pi^{(2,0)}+ 
\Pi^{(1,1)}\Pi^{(1,0)}
\label{eq:polepos} \\
&&+\, \Pi^{(3,0)}+ \Pi^{(2,1)}\Pi^{(1,0)}
+ \Pi^{(1,1)}\Pi^{(2,0)}+ [\Pi^{(1,1)}]^2\Pi^{(1,0)}+ 
\Pi^{(1,2)}[\Pi^{(1,0)}]^2+\ldots.
\nonumber\\[0.2cm]
R_{h\phi}^{-1} &=& 1 - \Pi^{(1,1)} - \Pi^{(2,1)} - 2 \Pi^{(1,2)}\Pi^{(1,0)}
\label{eq:residue} \\[0.1cm]
&&-\, \Pi^{(3,1)}-2  \Pi^{(2,2)}\Pi^{(1,0)}
-2  \Pi^{(1,2)}\Pi^{(2,0)}-
2\Pi^{(1,2)}\Pi^{(1,1)}\Pi^{(1,0)}- 
3 \Pi^{(1,3)}[\Pi^{(1,0)}]^2+\ldots.
\nonumber
\eeq
The first equation allows us to express the matching coefficient 
$\Delta$ [Eq.~(\ref{eq:Deltadef})] in terms of the hard part of 
the self-energy. Note that even though only the hard part of $\Pi$ 
appears in \eqref{eq:polepos}, ${\bar s}$ is the physical pole
location. In other words, there is no difference  whether we use the
full self-energy or only its hard part in \eqref{eq:polepos}, since 
$\Delta$ receives no soft contributions as discussed above.
The situation is different for the residue, since its hard
part, as defined in \eqref{eq:residue}, does not in general coincide
with the residue of the full propagator $R_\phi$ that appears in the
LSZ reduction formula and in \eqref{eq:propagator}. $R_\phi$ is given
by an equation similar to \eqref{eq:residue} where the hard part is
replaced by the full self-energy. The relation reads 
$R_\phi=R_{h\phi} R_{\rm eff\phi}$. The residues $R_\phi$ and 
$R_{\rm eff\phi}$ in the full and effective theory are IR divergent 
and depend on the IR regulator, but the hard part of the residue 
is independent of this regularization. However, if dimensional
regularization is adopted for the IR singularities, $R_{\rm
  eff\phi}=1$ (since all loops in the effective theory vanish), and 
the full and hard residues coincide. 

We now turn to the explicit calculation of the matching coefficient 
\be
\Delta \equiv \sum_i  \Delta^{(i)} = 
 \hM\, \Pi^{(1,0)} + 
 \hM  \left(\Pi^{(2,0)}+\Pi^{(1,1)}\Pi^{(1,0)} \right) + \ldots
\label{eq:Deltaexp}
\ee
with $\Delta^{(i)}=\cO{\alpha^i}$. We need $\Pi^{(1,0)}$ for the
calculation of the matching coefficient at leading order, while we
need $\Pi^{(2,0)}$ and $\Pi^{(1,1)}$ at NLO. 
Defining 
\be
\label{eq:adef}
a_g\equiv \frac{\alpha_g}{4\pi},\quad 
a_y\equiv \frac{\alpha_y}{4\pi},\quad 
a_\lambda\equiv \frac{\lambda}{16\pi^2},
\ee
we find  
\begin{eqnarray} 
  \Pi^{(1,0)} &=& a_y \Mmue
  \left(- \frac{2}{\eps} - 4 - 8 \eps + \frac{7\pi^2\,\eps}{6}
         - 2 i \pi - 4 i \pi \eps \right) 
\label{eq:Pi10} \\
  &+&  a_g \Mmue 
  \left(\frac{3}{\eps} + 7 + 15 \eps +\frac{\pi^2\,\eps}{4} \right) 
+ \delta_M^{(1)} - \delta_\phi^{(1)},
\nonumber \\
\Pi^{(1,1)} &=& a_y \Mmue 
   \left(-\frac{2}{\eps} -2 +\frac{7\pi^2\eps}{6} - 4 \eps 
         - 2 i \pi - 2 i \pi \eps \right) - \delta_\phi^{(1)},
\label{eq:Pi11}
\end{eqnarray}
where $\delta_M^{(1)}$ and $\delta_\phi^{(1)}$ refer to the 
order $\alpha$ part of the counterterms given in the Appendix. The 
order $\epsilon$ terms will be needed only in (\ref{eq:ctterm2}) below. 

The two-loop self-energy has
contributions proportional to $\alpha_y \alpha_g$ (4 diagrams),
$\alpha_y^2$ (2 diagrams) and $\alpha_g^2$ (10 diagrams), 
not counting  counterterm diagrams, as well as a contribution 
proportional to $\alpha_\lambda$, which by assumption is of the 
same order. The result
reads 
\begin{eqnarray}
  \Pi^{(2,0)} &=& 
  \Mmutwoe\, a_ya_g \,
  \Bigg[
  - \frac{3}{\eps^2} - \frac{17}{2 \eps} 
  + \frac{17\pi^2}{2} - \frac{99}{4} - 24 \zeta(3) 
  - \frac{6 i \pi}{\eps} - 39 i \pi 
  + \frac{8 i \pi^3}{3}
  \Bigg] \nn\\
  &+&
  \Mmutwoe\, a_y^2\, 
  \Bigg[ 
  - \frac{3}{\eps^2} - \frac{5}{2 \eps} 
  - \frac{11\pi^2}{2} + \frac{57}{4} 
  + \frac{2 i \pi}{\eps} + 5 i \pi  \Bigg] \nn\\
  &+&
  \Mmutwoe\, a_g^2\, 
  \Bigg[ 
  \frac{8}{\eps^2} + \frac{7}{2 \eps} 
  + \frac{44\pi^2}{3} - \frac{149}{4} + 24 \zeta(3)
  - 16 \pi^2 \ln2 \Bigg] \nonumber \\
  &-&  a_\lambda \Mmue
  \left(\frac{1}{\eps} + 1\right)
  + \Pi^{(2)}_{\rm ct}\>
\label{eq:Pi20}
\end{eqnarray}
with the counterterms given by
\beq
  \Pi^{(2)}_{\rm ct} &=& 
  \left(\delta_M^{(2)}-\delta_\phi^{(2)}\right) +
  \left( 2\delta_y^{(1)}-\delta_\psi^{(1)}
         -\delta_\chi^{(1)}\right) \Pi_y^{(1,0)}
  \nn\\
   &+& 
  \left( 2\delta_g^{\prime\, (1)}-\delta_A^{(1)}- \delta_\phi^{(1)}  
        \right) \Pi_g^{(1,0)}  \nn\\ 
   &+& 
  \left(\delta_M^{(1)}-\delta_\phi^{(1)}\right) 
  \hM^2 \frac{\partial\Pi_g^{(1,0)}(\hM^2,P^2)}{\partial \hM^2}
     \Bigg|_{P^2=\hM^2}
  + \delta_\lambda^{(2)}\, \int [dk]\,\frac{i}{k^2-\hat M^2}.
\label{eq:ctterm2}  
\eeq
In \eqref{eq:ctterm2} we indicated the power of $\alpha$ in the
counterterms and denoted by $\Pi_y^{(1,0)}$ and $\Pi_g^{(1,0)}$ 
the contribution to the one-loop self-energy
proportional to $\alpha_y$ and  $\alpha_g$ 
respectively, not including counterterms. We adopt the 
$\overline{\rm MS}$ renormalization scheme and refer to the 
Appendix for the explicit expressions of the counterterms 
appearing in (\ref{eq:ctterm2}). The two-loop integrals 
have been computed with standard techniques. We checked 
some of our results with recurrence relations from  
\cite{2loop/recurrence}, and two-loop master integrals from 
\cite{2loop/master}. Massive one-loop integrals (needed here and later) 
were in part computed with the method described in \cite{davydychev}. 

The matching coefficients $\Delta^{(1)}$  and $\Delta^{(2)}$ 
defined in \eqref{eq:Deltaexp}
read in the $\MSbar$-scheme 
\begin{eqnarray}
\frac{\Delta^{(1)}}{\hat M} &=& a_g \left(- 3\lnM +7\right) + a_y \left(
      2\lnM -4 - 2\, i\pi \right)
\label{eq:Delta1} \\
\frac{\Delta^{(2)}}{\hat M} &=&  a_g^2
\left(8\, \lnsqM  + \frac{16}{3} \lnM - \frac{193}{4} + \frac{40 \pi^2}{3} 
      - 16 \pi^2 \log(2) + 24 \zeta(3) \right) 
\nonumber \\
&+& a_y^2
\left( \lnsqM  - \big(11+10\,i\pi\big)\, \lnM  + \frac{89}{4} - 
  \frac{23\pi^2}{3} + 13\, i \pi \right)
\nonumber \\
&+& a_g a_y
\left(  - 9 \lnsqM + \big(31+ 12\, i\pi \big)\lnM - \frac{115}{4} + 
       5 \pi^2 - 24 \zeta(3)
       - 41\, i \pi + \frac{8\, i \pi^3}{3} \right)
\nonumber \\
&+& a_\lambda\left( \lnM -1\right).
\label{eq:Delta2}
\end{eqnarray}
We have computed these coefficients in an arbitrary covariant 
gauge and checked their gauge-parameter independence.  
Note that $\Pi^{(1,1)}$ and $\Pi^{(2,0)}$ in (\ref{eq:Pi11}) and
(\ref{eq:Pi20}) have $1/\epsilon$ infrared poles after $\MSbar$ 
renormalization, which must be associated 
with the factorization of the hard and soft contributions. 
For $\bar s$ and, therefore, for the matching coefficient
$\Delta$ these singularities cancel as they should. 
On the other hand, the residue at the pole, 
$R_\phi$, has an infrared singularity in the $\MSbar$-scheme. 
(In the on-shell scheme $R_\phi=1$ by definition, but in this scheme 
the field renormalization factor $Z_\phi$ exhibits the 
infrared singularity.)

\subsubsection{Including soft photons and fermions}
\label{sec:photferm}

To complete the soft Lagrangian ${\cal L}_{\rm HSET}$ for the NLO 
computation the bilinear terms of the unstable
field, ${\cal L}_{\phi\phi}$, have to be supplemented by the kinetic
terms of the soft photon and fermions. Due to gauge invariance, these
terms will also contain the interaction terms of a soft photon with a
soft charged fermion. 

The leading operators scale with $\delta^4$ and are given by the
gauge-invariant kinetic terms 
$\bar\psi_s i\!\!\not\!\!\!D_s \psi_s, \ \bar\chi_s i\!\!\!\not\!\!\partial
\chi_s$ and $-1/4\, F_{s\mu\nu}F_s^{\mu \nu}$. By convention the 
kinetic terms are canonically normalized, and their coefficients do
not receive corrections from hard loops. The effective Lagrangian 
contains an infinite number of higher-dimension operators 
such as $\bar\psi_s \psi_s \bar\chi_s \chi_s$ (which scales with 
$\delta^6$). At NLO we need the operators that scale with
$\delta^5$. However, since we assume weak coupling with 
$g\sim \delta^{1/2}$, at NLO we only need the operators that 
have non-zero coefficients as $g\to 0$. There are no such operators, 
so we conclude that at NLO the only relevant soft photon-fermion 
interaction is contained in $\bar\psi_s i\!\!\not\!\!\!D_s \psi_s$. 
No further calculation is required to match ${\cal L}_{\rm HSET}$ 
at NLO. 

Putting everything together we obtain for ${\cal L}_{\rm HSET}$
including all terms needed for the NLO line shape in the 
weak-coupling limit
\beq
{\cal L}_{\rm HSET} &=&  2 \hM  \phi_v^\dagger\, 
        \left( i v \cdot D_s - \frac{\Delta^{(1)}}{2} \right) \phi_v  +  
 2 \hM  \phi_v^\dagger\,
        \left( \frac{(i D_{s\top})^2}{2\hM} +
               \frac{[\Delta^{(1)}]^2}{8\hM} -
               \frac{\Delta^{(2)}}{2} \right) \phi_v 
\nonumber  \\
&-&\frac{1}{4}\,F_{s\mu\nu} F_s^{\mu\nu}
+\bar\psi_s i\!\not\!\!D_s \psi_s
+\bar\chi_s i\!\not\!\partial \chi_s
\label{heavyNLO}
\eeq
with $\Delta^{(1)}$ and $\Delta^{(2)}$ given in (\ref{eq:Delta1})
and (\ref{eq:Delta2}) respectively. The first term of \eqref{heavyNLO}
gives rise to the unstable particle propagator used in the
calculation of the leading-order amplitude \eqref{TLO}. All other
terms are needed at next-to-leading order only. Note that in writing 
the Lagrangian (\ref{heavyNLO}) we have expanded the 
matching coefficient $\Delta$ in powers of $\alpha$, so that 
the unstable particle 
propagator strictly counts as $1/\delta \sim 1/\alpha$. 
Alternatively, we could sum higher-order corrections to $\Delta$ 
back into the propagator, although this would not correspond 
to a strict expansion of the scattering amplitude in 
$\delta$ and $\alpha$. We will investigate the effect of including
additional terms in the propagator in the numerical analysis below. 

\subsection{\boldmath Matching ${\cal L}_{\rm SCET}$}
\label{sec:scet}

The leading Lagrangian for the collinear fields has been given 
in (\ref{scetLO}). In the position space formulation of soft-collinear
effective theory the Lagrangian has been worked out to order 
$\delta$ in \cite{BCDF}. (We note again that what is 
called ``soft'' here has been called ``ultrasoft'' in  \cite{BCDF}.)
When expanded systematically in powers of 
$\delta^{1/2}$, the interaction vertices are rather complicated. 
Our task of writing down all the terms needed for the NLO computation
of the line shape is again simplified by the fact that we adopt the
weak-coupling limit $g\sim\delta^{1/2}$. Inspection of the 
expressions in \cite{BCDF} shows that once again there are no relevant
operators at order $\delta^{9/2}$ and $\delta^5$ that survive as 
$g\to 0$. All the vertices needed for collinear loop corrections to
the LO line shape are contained in the leading-order Lagrangian 
\be
\label{scetterm}
\bar{\psi}_c \left(i n_- D +  i \Slash{D}_{\perp c}
\frac{1}{i n_+ D_{c}+i\epsilon}\, i\Slash{D}_{\perp c} \right)
\frac{\slash{n}_+}{2} \, \psi_c. 
\ee
The Feynman rules for the three-point vertices 
$\bar\psi_c \psi_c A_c$ and $\bar\psi_c \psi_c A_s$ can be read off 
from this expression. We may note that 
the inverse covariant derivative can be written in
terms of Wilson lines
\be
 \label{eq:WilsonProp}
  (i n_+ D_c + i\epsilon)^{-1}  =
  W_c \,(i n_+ \partial+i\epsilon)^{-1} \,W_c^\dagger,
\ee
where $W_c$ is 
\be
\label{eq:WilsonDef}
  W_c(x) = \exp\left(ig\int_{-\infty}^0 \!ds \,n_+ A_c(x+s
    n_+)\right). 
\ee 
The term (\ref{scetterm}) therefore contains vertices with any
number of $A_c$ fields, which are all leading power in $\delta$, but
suppressed by gauge coupling factors $g$, so they are
not needed for the computations of the NLO line shape.

Interactions with soft photons are contained in 
$\bar{\psi}_c \,i n_- D \,\psi_c$. In \cite{BCDF} collinear-soft
interaction vertices are light-cone multipole-expanded, and 
the covariant derivative reduces to 
$i\nm D = i\nm\partial + g \nm A_c(x)+g\nm A_{\rm
  s}(x_-)$ with $x_- = (n_+ x)\, n_-/2$ at leading power. 
This should be done here only for the coupling of soft fields 
to the fields describing collinear fluctuations, but not for 
the coupling to collinear fields with soft fluctuations, 
since only for the former fields is the transverse momentum 
large compared to the transverse momentum of soft fields. This 
affects the way momentum conservation is implemented at 
the corresponding interaction vertices. 

\subsection{Production (decay) vertex}
\label{sec:vertices}

The computation of the LO forward scattering amplitude (line shape) 
requires only the vertex $\phi_v\bar \psi_{c1}\chi_{c2}+\mbox{h.c.}$
at tree level. At NLO we therefore need the one-loop coefficient
function of this vertex, and production or/and decay vertices
suppressed by one factor of $\delta$ (without loop corrections). 

\subsubsection{Power-suppressed vertices}

The power-suppressed vertices with field content 
$\phi_v\bar \psi_{c1}\chi_{c2}$ can be obtained to order $\delta$ 
from the result for the heavy-to-light decay current 
in \cite{BCDF}. However, the result relevant for the 
forward scattering amplitude can be obtained in a simpler way, 
since we only wish to reproduce the tree $2\to2$ scattering 
diagram (\ref{TLO}) to better accuracy. 

For tree-level matching we can set $\Pi(s)=0$. Leaving out the
external spinors and setting $P=\hM v+k$, we obtain by expansion 
in the small momentum $k$
\begin{equation}
\label{treeexpand}
 iy\frac{i}{P^2-\hM^2}iy^* = 
 iy\frac{i}{2\hM v\cdot k}iy^* + 
 iy\frac{i}{2\hM v\cdot k}ik_\top^2\frac{i}{2\hM v\cdot k}iy^* +
 i \frac{y y^*}{4\hM^2} + {\cal O}(\delta).
\end{equation}
The first term on the right-hand side is accounted for by the 
leading-order term in ${\cal L}_{\rm HSET}$ together with the 
leading-order production (decay) vertex. The second term 
involves one insertion of the kinetic energy correction 
$(iD_{s\top}^2)$ in (\ref{heavyNLO}). (For head-on collisions 
this term vanishes at tree level, because $k_\top=0$.) The third 
term is not reproduced by any term in the Lagrangians we discussed 
so far. Since the intermediate unstable scalar propagator 
has been canceled by the expansion, we associate this term 
with the production-decay operator
\begin{equation}
\label{pdop}
T = \frac{yy^*}{4 \hM^2} \,(\bar\psi_{c1}\chi_{c2})
                (\bar\chi_{c2}\psi_{c1}).
\end{equation}

It should be noted that this operator contributes at tree level 
contrary to the off-shell scattering topologies in the full theory 
discussed in the 
previous section. When we distinguish collinear fields with collinear
fluctuations from those with 
soft fluctuations, the operator discussed here appears with field 
content $(\bar\psi_{n_-}\chi_{n_+})(\bar\chi_{n_+}\psi_{n_-})$, 
since each fermion pair describes a configuration with invariant mass 
close to $\hM^2$. This must be distinguished from the off-shell
configurations produced in the scattering of a collinear electron 
fluctuation off the neutrino. This scattering process is represented
at tree-level by a non-local four-fermion operator, but it does not
contribute at tree-level to the forward scattering amplitude 
in the effective theory,
resulting in a NNLO contribution. 

\subsubsection{Hard loop correction}

We now turn to the one-loop matching of the coefficient function 
$C$ of the production (decay) operator
\be
\label{jj}
 J(x) = y\,e^{-i \hM v x}\,
[\phi_v \bar \psi_{c1} W_{c1} \chi_{c2}](x) + \rm{h.c.} 
\ee
Here we have added the Wilson line $W_{c1}$ (see (\ref{eq:WilsonDef})) 
to the definition of the operator. This generates an infinite number 
of interaction vertices with additional $n_+ A_{c1}$ photons fields, 
which are not suppressed by powers of $\delta$, but all have the 
same short-distance coefficient $C$ because of gauge invariance 
\cite{Bauer:2000yr}. The Wilson line accounts for full-theory 
diagrams where a soft $\phi$-field emits one or several $c1$-collinear
photons and, thereby, becomes off-shell. Integrating out the hard 
internal $\phi$-field results in the Wilson line. 
At next-to-leading order we are only concerned with
the emission of at most one collinear photon, which must then 
couple to the external electron line attaching to the same 
production (decay) vertex from which the collinear photon is 
emitted. We also mention that in principle we should consider 
the emission of photons collinear with the neutrino from the 
electron or the heavy scalar, which puts the electron or scalar lines
off-shell. However, the effect cancels due to the
charge-neutrality of the neutrino. It is for this reason that 
we do not introduce a $c2$-collinear photon field in the effective 
theory. 

We begin  the one-loop computation of $C$ by specifying 
the matching prescription. In the underlying theory we 
compute the amputated, $\overline{\rm MS}$-renormalized, 
on-shell three-point function of an electron, 
neutrino and scalar field, multiplied by the LSZ residue
factors, and expand it to leading order in $\delta$. The corresponding
three-point function in the effective theory contains one insertion of
the operator  $y\,e^{-i \hM v x}\,
[\phi_v \bar \psi_{c1} W_{c1} \chi_{c2}](x)$. Denoting this
three-point function by $\Gamma_{\rm eff}$ and the former by 
$\Gamma$, the matching equation reads
\begin{equation}
\label{eq:match3gen}
\sqrt{R_\psi} \sqrt{R_\chi}\sqrt{R_\phi}\, \Gamma = 
C\,\varpi^{-1/2}\sqrt{R_{\rm eff \psi}} 
\sqrt{R_{\rm eff \chi}}\sqrt{R_{\rm eff \phi}}\, \Gamma_{\rm eff}
\end{equation}
Here the various $R$-factors denote the ultraviolet-finite 
LSZ residues in the $\overline{\rm MS}$ scheme given by the ratios of
the on-shell and $\overline{\rm MS}$ field renormalization 
constants. The factor $\varpi^{-1/2}$ defined in (\ref{eq:Phinorm}) 
appears because of the non-relativistic normalization of the field
$\phi_v$. Its origin can be understood most directly by comparing 
the scalar propagators in full and effective theory diagrams 
as done in (\ref{eq:effprop}). Note that the on-shell condition for
the scalar line implies that its momentum satisfies $P^2=\bar s =
\hM^2+\hM\Delta$, but in a perturbative matching calculation this 
condition must be fulfilled only to the appropriate order in 
$\alpha$ and $\delta$. For the one-loop matching of $C$ it is sufficient
to put $P^2=\hM^2$. 

The matching equation \eqref{eq:match3gen} is valid for any infrared
regulator, but becomes particularly simple in the case of 
dimensional regularization, where all loop diagrams in the effective
theory vanish on-shell. Then all the effective theory residues equal
1, $\Gamma_{\rm eff}$ equals its tree-level value and 
the matching equation becomes
\be
\label{eq:match3dr}
\sqrt{R_{h\psi}} \sqrt{R_{h\chi}} \sqrt{R_{h\phi}} \sqrt{\varpi} \
\Gamma_h = C\, \Gamma_{\rm tree} 
\ee
The full theory quantities must then be equal to their hard contribution 
in the sense of the expansion by regions, and since $\varpi$ can be 
expressed in terms of the short-distance coefficient $\Delta$, 
this equation says that we can obtain $C$ by directly computing the 
hard contributions in the full theory. 

The residue $R_{h\phi}$ can be obtained from
(\ref{eq:residue}) and (\ref{eq:Pi11}), and the residues $R_{h\psi}$ and
$R_{h\chi}$ can be computed in a similar way. The results are given in
the Appendix. There is only one one-loop diagram and one counterterm
diagram contributing to $\Gamma$ at next-to-leading order. Evaluating
the hard part of this one-loop diagram we obtain 
\begin{equation}
\Gamma_h = i y \left( 1+ a_g \Mmue \left( - \frac{1}{\eps^2}
  - \frac{1}{\eps} - 2 - \frac{\pi^2}{12} \right) + \delta_y^{(1)}
\right)
\label{eq:Gammah}
\end{equation}
We can now apply \eqref{eq:match3dr} to find the matching coefficient at
NLO. Together with
\be
\varpi = 1-\frac{\Delta^{(1)}}{2 \hM} + \cO{\alpha^2,\alpha\delta,\delta^2} 
\ee 
and the residues from the Appendix we obtain  $C = 1 + C^{(1)} + 
\cO{\alpha^2}$, where 
\begin{eqnarray} 
  C^{(1)} &=&
  a_y \Bigg[\log \frac{\hM^2}{\mu^2} -\frac{1}{4} -\frac{i\pi}{2} \Bigg] 
\nn \\ 
  &+& a_g \Bigg[-\frac{1}{\eps^2} + 
              \frac{1}{\eps} \left(\lnM-\frac{5}{2}\right)
              - \frac{1}{2} \ln^2\frac{\hM^2}{\mu^2} 
              +\frac{7}{4}\lnM - \frac{15}{4} -
              \frac{\pi^2}{12}\Bigg]. 
\label{C1final}
\end{eqnarray}
The poles are infrared poles related to the factorization of hard and 
soft contributions. The final result for $C^{(1)}$ is obtained 
by subtracting the pole part. (In the following $C^{(1)}$ refers to
this subtracted expression.) It is independent of
the gauge parameter $\xi$ as it should be, because the production
(decay) operator is gauge-invariant. The double logarithm in the coefficient
of $a_g$ is related to the infrared divergence in $\Gamma_h$. The
single logarithms on the other hand are also related to the scale 
dependence of the Yukawa coupling $y$.

\subsection{Forward scattering amplitude in the effective 
theory}

We are now in the position to compute the forward scattering amplitude
at next-to-leading order in the effective theory. 

\begin{figure}[t]
\vskip0.2cm
\begin{center}
\begin{picture}(250,80)(0,0) 
  \ArrowLine(20,40)(10,10)
  \ArrowLine(10,70)(20,40)
  \Line(20,39)(80,39)
  \Line(20,41)(80,41)
  \GCirc(20,40){2}{0}
  \GCirc(80,40){2}{0}
  \BCirc(50,40){4} \Text(50,40)[c]{$\times$}
  \ArrowLine(90,10)(80,40)
  \ArrowLine(80,40)(90,70)
  \SetOffset(150,0)
  \ArrowLine(20,40)(10,10)
  \ArrowLine(10,70)(20,40)
  \Line(20,39)(80,39)
  \Line(20,41)(80,41)
  \BCirc(20,40){4} \Text(20,40)[c]{$\times$}
  \GCirc(80,40){2}{0}
  \ArrowLine(90,10)(80,40)
  \ArrowLine(80,40)(90,70)
\end{picture}
\end{center}
\vskip-0.3cm
\centerline{\parbox{14cm}{\caption{\label{fig:nlohard}\small  
Hard contributions to the next-to-leading order amplitude. Left:
Insertion of $[\Delta^{(1)}]^2/4$ and $-\hM\Delta^{(2)}$.  Right:
Insertion of $C^{(1)}$, the symmetric diagram is
understood. }}}\end{figure}
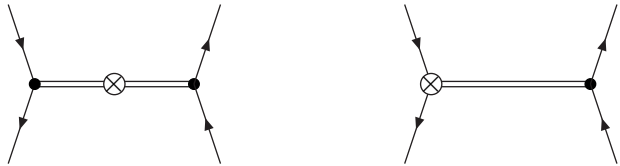

The first class of contributions consists of
diagrams with the topology of the tree diagram (\ref{TLO}), 
but with corrections to the propagator and the production (decay) 
vertex as shown in Figure~\ref{fig:nlohard}. As can be seen from
\eqref{heavyNLO}, the relevant terms for the left diagram of
Figure~\ref{fig:nlohard} are $[\Delta^{(1)}]^2/4$ and $-\hM
\Delta^{(2)}$. The remaining term $(i D_{s\top})^2$ does not
lead to a contribution at NLO. First for head-on collisions 
the scalar line has $k_{\top}=0$. Secondly, the photon vertices 
from this term contribute only at NNLO for weak coupling $g$. 
The other diagrams in this class are obtained from the hard
corrections to the Wilson coefficient $C$ of the production (decay) 
vertex. (The production and decay operator have the 
same complex coefficient $C$, and not the complex conjugates to 
each other, see (\ref{jj}).)  
There are two such diagrams, one of which is displayed in
Figure~\ref{fig:nlohard}. Adding up these three diagrams we obtain
\beq
i\,{\cal T}^{(1)}_{h} = i\,{\cal T}^{(0)} \!\!&\times& \!\!
\Bigg[ 2\, C^{(1)} - \frac{[\Delta^{(1)}]^2}{8\cD\hM} 
        + \frac{\Delta^{(2)}}{2\cD}  \ \Bigg]
\label{eq:T1h}
\eeq
with
\be
\cD\equiv \sqrt{s} - \hM - \frac{\Delta^{(1)}}{2}.
\ee
All these diagrams correspond to corrections
due to hard loops and, thus, can be identified with the factorizable
corrections \cite{effective}. These corrections are separately gauge
invariant since so are the matching coefficients. In addition 
there is the contribution 
from the production-decay operator $T$ in (\ref{pdop}), 
which contributes to $i\,{\cal T}^{(1)}_{h}$ the term
\beq
\nonumber \\[-0.1cm]
  \unitlength .7pt\SetScale{0.8}
  \begin{picture}(100,30)(0,30)
    \ArrowLine(30,30)(0,0)
    \ArrowLine(0,60)(30,30)
    \GCirc(30,30){2}{0}
    \ArrowLine(60,0)(30,30)
    \ArrowLine(30,30)(60,60)
  \end{picture}
= \frac{i y y^* s}{8 \hM^2} 
= i \,{\cal T}^{(0)} \times \left[-\frac{\cD}{2\hM}\right]. 
\label{eq:fourferm} 
\\ \nonumber
\eeq 

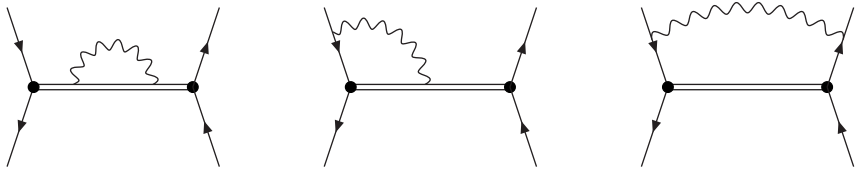
\begin{figure}[t]
\vskip0.2cm
\begin{center}
\begin{picture}(350,80)(0,0) 
  \ArrowLine(20,40)(10,10)
  \ArrowLine(10,70)(20,40)
  \Line(20,39)(80,39)
  \Line(20,41)(80,41)
  \GCirc(20,40){2}{0}
  \GCirc(80,40){2}{0}
  \PhotonArc(50,41)(15,0,180){2}{7}
  \ArrowLine(90,10)(80,40)
  \ArrowLine(80,40)(90,70)
  \SetOffset(120,0)
  \ArrowLine(20,40)(10,10)
  \ArrowLine(10,70)(20,40)
  \Line(20,39)(80,39)
  \Line(20,41)(80,41)
  \GCirc(20,40){2}{0}
  \GCirc(80,40){2}{0}
  \PhotonArc(25,41)(23,0,120.5){2}{7}
  \ArrowLine(90,10)(80,40)
  \ArrowLine(80,40)(90,70)
  \SetOffset(240,0)
  \ArrowLine(20,40)(10,10)
  \ArrowLine(10,70)(20,40)
  \Line(20,39)(80,39)
  \Line(20,41)(80,41)
  \GCirc(20,40){2}{0}
  \GCirc(80,40){2}{0}
  \PhotonArc(50,10)(60,53,127){2}{9.5}
  \ArrowLine(90,10)(80,40)
  \ArrowLine(80,40)(90,70)
\end{picture}
\end{center}
\vskip-0.4cm
\centerline{\parbox{14cm}{\caption{\label{fig:nlosoft}\small  
Soft contributions to the next-to-leading order amplitude. A diagram
with a soft correction at the decay vertex is
understood. }}}\end{figure}

We now turn to the one-loop corrections in the effective theory, which
are due to soft or collinear photons. The loop provides a factor of 
$\alpha$, so at NLO in a combined expansion in $\alpha$ and $\delta$ 
we can use the leading power soft and collinear photon vertices. 
We begin with the soft correction.  
The coupling of the soft photon to the $\phi_v$-field
is given by the covariant derivative in the 
first term in ${\cal L}_{\rm HSET}$ \eqref{heavyNLO}, 
whereas the coupling to the collinear electron can be obtained from
the first term in ${\cal L}_{\rm SCET}$ \eqref{scetLO}. With
dimensional infrared regularization the residue factors equal 1, 
which leaves the diagrams shown in Figure~\ref{fig:nlosoft}.
For the forward scattering amplitude the
box diagram (third diagram in Figure~\ref{fig:nlosoft}) 
vanishes in Feynman gauge, because the photon-coupling to the 
collinear electron is proportional to $n_-^\mu$. Thus the amplitude is
proportional to $n_-^2=0$.
Computing the remaining diagrams of Figure~\ref{fig:nlosoft} we obtain  
\beq
i\,{\cal T}^{(1)}_{s} = i\,{\cal T}^{(0)} \!\!&\times& \!\! a_g\, 
\left(\frac{-2\cD}{\mu}\right)^{-2\eps} 
\Bigg[\, \frac{2}{\eps^2}+\frac{2}{\eps}
+ 4 + \frac{5\pi^2}{6} \Bigg].
\label{eq:T1s}
\eeq
The final result for the soft amplitude is the above expression 
with the pole terms subtracted after expanding the prefactor 
$(-2\cD/\mu)^{-2\eps}$. 
The soft part of the amplitude does not contain logarithms
of the hard scale as expected. We also note that 
the soft amplitude is gauge-invariant, because it is computed 
with a gauge-invariant effective Lagrangian. 

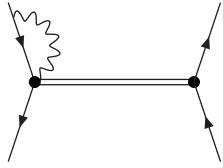
\begin{figure}[t]
\vskip0.2cm
\begin{center}
\begin{picture}(350,80)(0,0) 
  \SetOffset(120,0)
  \ArrowLine(20,40)(10,10)
  \ArrowLine(10,70)(20,40)
  \Line(20,39)(80,39)
  \Line(20,41)(80,41)
  \PhotonArc(15.667,53)(12,-71,107){2}{6.5}
  \GCirc(20,40){2}{0}
  \GCirc(80,40){2}{0}
  \ArrowLine(90,10)(80,40)
  \ArrowLine(80,40)(90,70)
\end{picture}
\end{center}
\vskip-0.4cm
\centerline{\parbox{14cm}{\caption{\label{fig:nlocol}\small  
Contributions to the next-to-leading order amplitude involving
collinear photons.}}}
\end{figure}

The only collinear photon correction at NLO comes from the 
diagram shown in Figure~\ref{fig:nlocol} (and the corresponding 
symmetric diagram), where the collinear photon coupling at the 
production (decay) vertex is derived from the Wilson line in 
(\ref{jj}). However, for an on-shell electron the integral is 
scaleless and vanishes in dimensional regularization.

Combining \eqref{eq:T1h}, \eqref{eq:fourferm}, (\ref{eq:T1s}) with 
the result for the matching coefficient
$C^{(1)}$ in (\ref{C1final}) we obtain for the forward scattering 
amplitude at next-to-leading order 
\beq
i\,{\cal T}^{(1)} = i\,{\cal T}^{(0)} &\times& \! \Bigg[\ 
a_g \left( 3 +4\, \log\frac{-2\hM \cD}{\hM^2} \right)
    \left(- \frac{1}{\eps} +\log\frac{-2\hM\cD}{\mu^2} \right)
\nn \\
&& +\ a_g  \left( 
             - 7\, \log\frac{-2\hM \cD}{\hM^2} 
             - \frac{3}{2} \log\frac{\hM^2}{\mu^2}
              -\frac{7}{2} + \frac{2\pi^2}{3} \right) 
\nn \\
&& +\ a_y \left( 2\, \log\frac{\hM^2}{\mu^2} 
                 -\frac{1}{2} -i\pi \right) 
   - \frac{[\Delta^{(1)}]^2}{8\cD\hM} 
   + \frac{\Delta^{(2)}}{2\cD} - \frac{\cD}{2\hM} \ \Bigg].
\label{eq:T1}
\eeq
Rather than inserting the subtracted expressions for ${\cal
  T}^{(1)}_s$ and $C^{(1)}$ we have used here the unsubtracted ones 
to display the structure of singularities. The double poles in 
$1/\eps$ have canceled between the hard and soft corrections, but
there are $1/\eps$ poles left over from ${\cal T}^{(1)}$, which must
be associated with the collinear singularity due to initial state
radiation from the massless electron. To verify that these singularities have
the expected structure we compute 
\beq
[i\,{\cal T}^{(1)}_{\rm sing}/s] =
-\frac{\alpha_g}{2\pi} \int_0^1 dx\, 
\frac{1}{\eps} \,P(x)\, [i\,{\cal T}^{(0)}(xs)/(x s)]
\eeq
where $i\,{\cal T}^{(0)}(x s)$ is the leading-order forward scattering
amplitude given in 
(\ref{TLO}), but with $s$ replaced by $x s$. ($x$ is the momentum 
fraction of the electron in the electron.) Furthermore 
\begin{equation}
P(x) = \frac{2}{\,\,[1-x]_+}+\frac{3}{2}\,\delta(1-x)
\end{equation}
is the soft limit of the $e\to e$ Altarelli-Parisi splitting 
function, and we took into account that the incoming neutrino does not
radiate photons. Simplifying the splitting function 
to the soft limit is justified, because values of $x$ not near one 
correspond to hard-collinear radiation, and this is a NNLO effect 
(which is part of the production-decay operator in the right panel of 
Figure~\ref{fig:skeleton}). Similarly the $x$-integral could be
restricted to an interval close to one. Equivalently, we perform the 
$x$-integration from zero to one and expand the result around 
$\sqrt{s}=\hM$ to obtain 
\beq
i\,{\cal T}^{(1)}_{\rm sing} = 
i\,{\cal T}^{(0)} \times \left(-\frac{a_g}{\eps}
\right)\left(3+ 4 \log\frac{-2\hM \cD}{\hM^2} \right),
\label{eq:singsig1}
\eeq
which agrees with the pole part of \eqref{eq:T1}. This 
confirms that the cross section is finite once the
initial-state singularities have been factored into the electron
distribution function. 

Subtracting the initial-state singularities minimally according 
to our convention for the electron distribution function, we 
obtain
\beq
i\,{\cal T}^{(1)} = i\,{\cal T}^{(0)} &\times& \! \Bigg[\ 
a_g \left( 3 \log\frac{-2\hM\cD}{\nu^2}  
   + 4\, \log\frac{-2\hM \cD}{\hM^2} \log\frac{-2\hM\cD}{\nu^2}
    \right.
\nn \\
&&  \left. \qquad -\ 7 \log\frac{-2\hM \cD}{\hM^2} 
           - \frac{3}{2} \log\frac{\hM^2}{\mu^2}
           -\frac{7}{2} + \frac{2\pi^2}{3} \right) 
\nn \\
&& +\ a_y \left( 2 \log\frac{\hM^2}{\mu^2} 
                 -\frac{1}{2} -i\pi \right) 
   - \frac{[\Delta^{(1)}]^2}{8\cD\hM} 
   + \frac{\Delta^{(2)}}{2\cD} - \frac{\cD}{2\hM} \ \Bigg]
\label{eq:T1finite}
\eeq
as the final result for the next-to-leading order contribution to 
the forward scattering amplitude. The correction to the 
inclusive line shape is given by the imaginary part of this 
expression. We have introduced the scale $\nu$ to distinguish the 
scale of the electron distribution function from the renormalization
scale in the masses and couplings.  Referring to
\eqref{fullLO} we note that $\nu$ should be taken of 
order $\hM \delta$ to make the initial-state collinear 
logarithms small. 

We performed several further checks of this result. First we find that 
the logarithms of $\mu$ cancel the scale dependence of $\hM$ and 
$yy^*$ in $i\,{\cal T}^{(0)}$ to NLO. Second we expand $1/s\times 
\mbox{Im}\,({\cal T}^{(0)}+{\cal T}^{(1)})$ to order $\alpha^2$ 
and compare this to the expansion of the fixed-order cross section 
(\ref{fullLO}) to order $\alpha^2/\delta$. The 
two expansions coincide in the terms $\alpha^2/\delta^2$, 
$\alpha^2/\delta$, which they have in common. 
Finally, we have also computed the 
$\delta$ expansion of the one-loop forward scattering amplitude 
directly in the full theory.  
This checks all the terms in (\ref{eq:T1finite}) except for 
those involving $\Delta^{(2)}$ and $[\Delta^{(1)}]^2$, which come from
two-loop diagrams in the full theory. 

The final result (\ref{eq:T1finite}) is given in the $\overline{\rm MS}$ 
scheme. We briefly discuss how to translate this result to the pole  
renormalization scheme, where the mass $M$ is defined by 
$\bar s=M^2-i M\Gamma$ and the residue of the renormalized 
$\phi$-propagator is $R_\phi=1$. All other renormalization conventions 
remain unchanged. From the definition of $\Delta$ in (\ref{eq:Deltadef}) 
we obtain in the pole scheme $\mbox{Re}\,\Delta_p=0$ and 
\begin{equation}
\label{imdelta}
\mbox{Im}\,\frac{\Delta_p}{M} = \frac{\mbox{Im}\,(\Delta/\hat M)}
{1+\mbox{Re}\,(\Delta/\hat M)}
\end{equation}
with $\Delta$ and $\hM$ referring to the $\overline{\rm MS}$ (or, in fact, 
any other) scheme. Using (\ref{eq:Delta1}), (\ref{eq:Delta2}) 
we obtain 
\begin{eqnarray}
\frac{\Delta^{(1)}_p}{M} &=& - 2\, i\pi a_y,
\label{eq:Delta1pole} \\
\frac{\Delta^{(2)}_p}{M} &=&  i\pi a_y^2
\left(- 6 \ln\frac{M^2}{\mu^2} + 5\right)
+i\pi a_g a_y
\left(6 \ln\frac{M^2}{\mu^2} - 27 + \frac{8 \pi^2}{3} \right).
\label{eq:Delta2pole}
\end{eqnarray}
To convert the matching coefficient $C$ to the pole scheme, 
we observe that $R_\phi Z_\phi$ is scheme-independent, so the only 
modification comes from the normalization factor $\varpi$, 
resulting in 
\begin{equation}
\label{cpole}
C_p = \left(\frac{\varpi_p}{\varpi}\right)^{1/2} C,\qquad 
2 C^{(1)}_p = 2 C^{(1)} + \frac{\mbox{Re}\,\Delta^{(1)}}{2\hM}.
\end{equation}
This is exactly what is needed to render the line shape 
scheme-independent to leading order in $\delta$ and to all 
orders in $\alpha$. Using (\ref{imdelta}) and (\ref{cpole}) 
together with $M^2=\hM^2+\hM\,\mbox{Re}\,\Delta$ we 
find $C_p^2/M = C^2/\hM$, so 
\begin{equation}
\frac{C_p^2}{M\Big(\sqrt{s}-\sqrt{M^2+M\Delta_p}\Big)} = 
\frac{C^2}{\hM\Big(\sqrt{s}-\sqrt{\hM^2+\hM\Delta}\Big)}
\end{equation}
exactly.

\subsection{Next-to-leading order line shape}
\label{sec:corr_ls}

We compute the line shape at NLO in the
$\overline{\rm MS}$ renormalization scheme by taking the imaginary
part of $({\cal T}^{(0)}+{\cal T}^{(1)})/s$. The result 
depends on the renormalization scale $\mu$ and rather strongly on 
the collinear factorization scale $\nu$.  This 
dependence would be canceled (up to the order at which we
perform the calculation), if we folded our cross section with the
electron distribution function. Since this is well understood 
and we are interested in displaying the next-to-leading 
order effects intrinsic to the line-shape calculation, 
we present the partonic cross section, and adopt  $\mu=
M$ and $\nu=|2\cD|$ unless otherwise stated, since this choice 
makes the collinear logarithms unimportant. 

\begin{figure}[t]
  \unitlength 1cm
  \begin{center}
  \begin{picture}(13.4,6)
  \put(-1.0,-0.2){\includegraphics{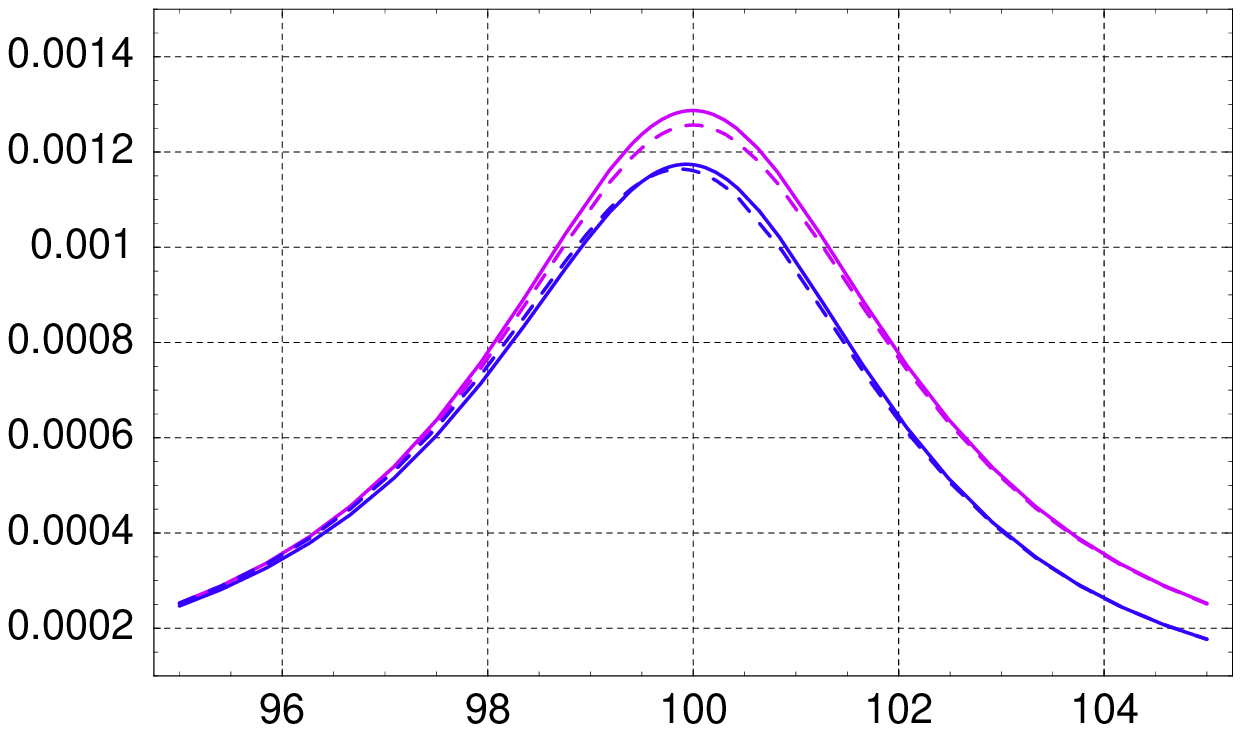}}
  \put(-0.5,-6){\includegraphics{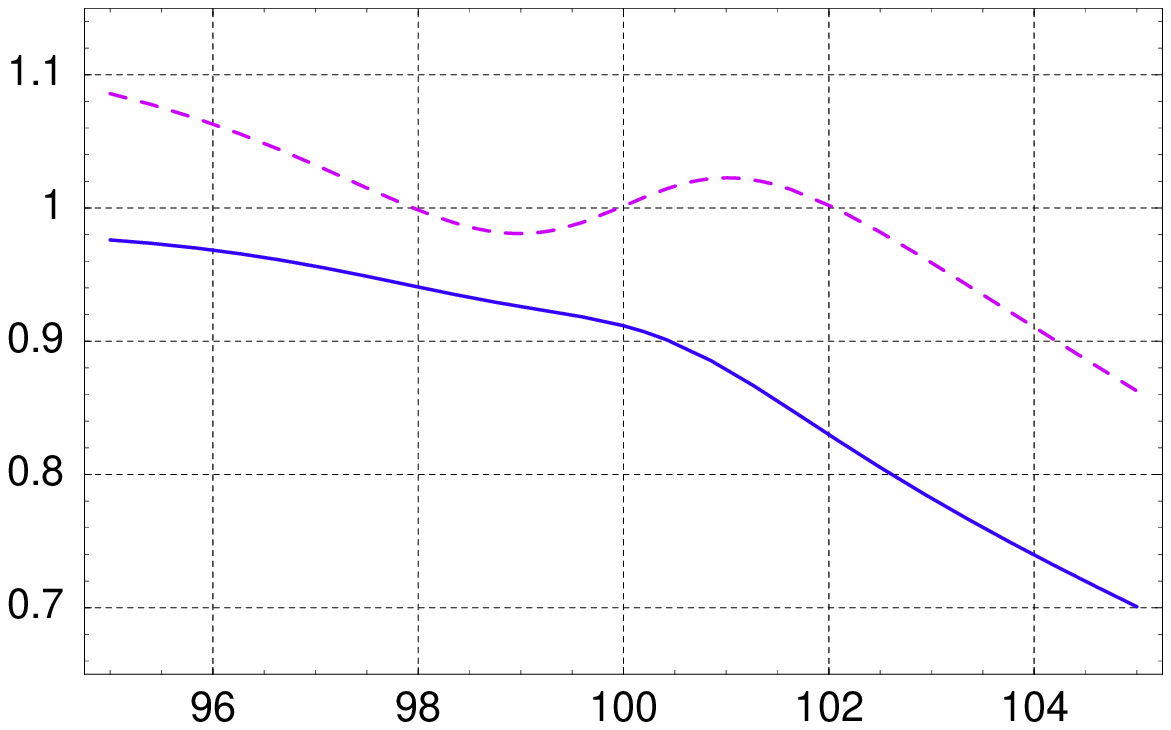}}
  \end{picture} 
  \end{center}
  \vskip5.6cm
  \centerline{\parbox{14cm}{\caption{\label{fig:nloLS} \small
  Upper panel: line shape (in $\mbox{GeV}^{-2}$) in the 
  $\overline{\rm MS}$ scheme (solid) and pole scheme
  (dashed) at LO (upper magenta/light grey curves) and NLO
  (lower blue/dark grey curves) as a function of the center-of-mass 
  energy in GeV. 
  Lower panel: The ratio of the NLO to the LO line shape in 
  the $\overline{\rm MS}$ scheme (solid blue/dark grey curve) 
  and the ratio of the NLO  $\overline{\rm MS}$ line shape to a 
  Breit-Wigner fit (dashed magenta/light grey curve).}}}
\end{figure}

In order to illustrate the numerics we use $a_g=0.1/(4\pi)$,
$a_y=0.1/(4\pi)$ and $a_\lambda=(0.1)^2/(4\pi)^2$ and assume the pole
mass $M=100$~GeV. For the plots in the $\overline{\rm MS}$ scheme
we have to convert the pole mass to the $\overline{\rm MS}$ mass
$\hM$. At LO we use the one-loop relation between the pole mass and
$\hM$ and obtain $\hM=98.8$~GeV, whereas at NLO we use the two-loop
relation and get $\hM=99.1$~GeV. The LO line shape has already been
shown in Figure~\ref{fig:line-shape}. In Figure~\ref{fig:nloLS} (upper
panel) we 
show the LO and the NLO line shape in the $\overline{\rm MS}$ scheme
(solid curves) and the pole scheme (dashed curves). The scheme
dependence is very small, but the next-to-leading order correction 
is around $-10\%$ near the peak, and can be up to $-30\%$ for 
center-of-mass energies between 95~GeV and 105~GeV . 
The ratio of NLO to LO is shown as the solid line in the 
lower panel of the figure. 

Contrary to the LO line-shape the NLO line shape does not any
longer have an exact (non-relativistic) Breit-Wigner shape. Performing
a fit of the NLO line shape ($\overline{\rm MS}$ scheme) 
to a Breit-Wigner curve in the range 
$95 \le \sqrt{s} \le 105$~GeV we find deviations up to $15\%$. 
The ratio of the NLO curve to the Breit-Wigner fit is shown
in as dashed line in Figure~\ref{fig:nloLS}. Moreover, the output 
mass parameter of the Breit-Wigner fit differs from the input 
$M=100\,$GeV by $-160\,$MeV. This illustrates that in realistic
situations the data should be fitted to the theoretically predicted
shape rather than a Breit-Wigner-type ansatz. 

\begin{figure}[t]
  \unitlength 1cm
  \begin{center}
  \begin{picture}(10,6)
  \put(-2.3,-0.3){\includegraphics{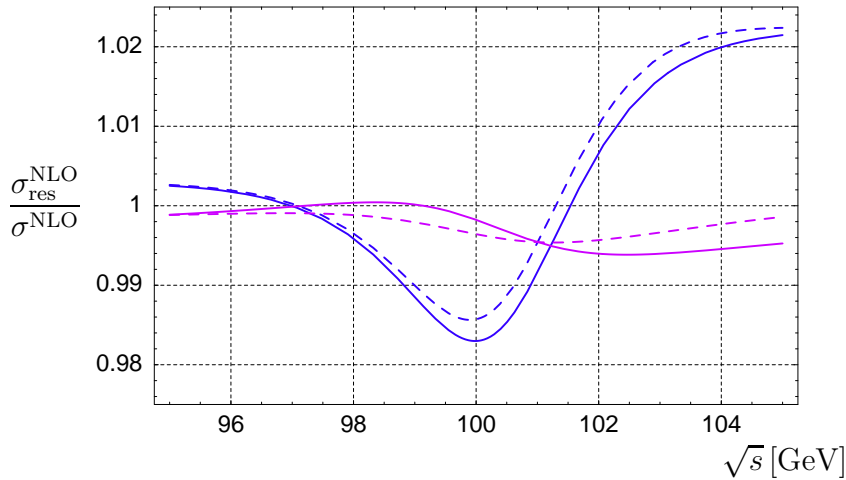}}
  \put(8.5,-0.6){$\sqrt{s}\,$[GeV]}
  \put(-1.0,2.8){$\displaystyle
               \frac{\sigma^{\rm NLO}_{\rm res}}{\sigma^{\rm NLO}}$}
  \end{picture} 
  \end{center}
  \vskip0.1cm
  \centerline{\parbox{14cm}{\caption{\label{fig:resD} \small
   The ratio of the line shape with powers of $\Delta^{(2)}$
   (and $\Delta^{(1)}$) resummed over the non-summed line shape in
   the pole (magenta/light grey lines) and $\overline{\rm MS}$ 
   scheme (blue/dark grey lines). The dashed  curves represent the
   line shape including
   $\Delta^{(2)}$ in the propagator. Resumming also all powers of 
   $\Delta^{(1)}$ results in the solid curves.}}}
\end{figure}

Up to now we have performed a strict expansion in the couplings and $\delta$
to obtain the NLO result. In particular, we only included
$\Delta^{(1)}$ in the propagator and treated the terms $\Delta^{(2)}$
as well as higher powers of $\Delta^{(1)}$ as interactions in a
perturbative expansion. In other words, we used $i/(2 \hM\cD)$ for the
propagator of the unstable field. A slightly different approach is to
include the effect of higher-order bilinear terms in the Lagrangian into the 
propagator, which sums these terms to all orders. This can be done in 
different versions. For example, instead of $\Delta^{(1)}$ we may 
use the full $\Delta$. At NLO, this implies using the 
propagator
\beq
\frac{i}{2 \hM (\cD -\Delta^{(2)}/2)}
\eeq
which absorbs the interaction term $\Delta^{(2)}/2\cD$ and 
sums all powers of $\Delta^{(2)}$ in the
forward scattering amplitude. The difference to the non-summed 
line shape is of course a NNLO effect. The ratio of summed 
to non-summed line shape at NLO is shown as the dashed curves 
in Figure~\ref{fig:resD} in the $\overline{\rm MS}$ scheme
(blue/dark grey line) and the pole scheme (magenta/light grey line). 
In computing the resummed line shape we have
adapted the factorization scale and used $\nu=|2\cD-\Delta^{(2)}|$. 
It can be seen from Figure~\ref{fig:resD} that the effect 
of resummation is less than 2\%, which indicates that the 
higher-order terms associated with $\Delta^{(2)}$ are not large. 
The effect is still smaller in the pole scheme. 

We can go one step further and also resum all terms 
associated with the expansion of~$\Delta$. In order to achieve this we
have to recall that we obtained the propagator by expanding $vk =
-\hM+\sqrt{\hM^2+\hM\Delta-k_{\top}^2}$ in $\Delta$ and $k_{\top}$ 
(see (\ref{eq:eqofmotion})). Expanding only $k_{\top}^2$ we obtain
the summed propagator, which at NLO reads 
\beq
\frac{i}{2 \hM (\sqrt{s}-
   \sqrt{\hM^2 + \hM (\Delta^{(1)}+\Delta^{(2)})})}.
\eeq
This also absorbs the interaction term $-[\Delta^{(1)}]^2/8 M \cD$ 
into the propagator. The ratio of corresponding summed 
to non-summed line shape at NLO is shown as the solid curves 
in Figure~\ref{fig:resD} in the $\overline{\rm MS}$ scheme
(blue/dark grey line) and the pole scheme (magenta/light grey line). 
(The factorization scale has been adapted to $\nu=2 
|(\sqrt{s}-\sqrt{M^2 + M(\Delta^{(1)}+\Delta^{(2)})})|$.) 
Again the effect of resummation is marginal compared to the size of 
the NLO correction.

\section{Beyond NLO \label{sec:beyondnlo}}

A crucial aspect of our formalism is that it allows to improve the
accuracy of the calculations by including systematically higher-order 
corrections.  After the explicit calculation of the line
shape at NLO, we now discuss how to go beyond NLO. We will outline the 
necessary calculations to obtain the line shape at NNLO in the effective 
theory approach. Then we compute explicitly a subset of corrections 
to illustrate once more how the formalism automatically maintains 
gauge invariance of the calculation. 

\subsection{Elements of the NNLO calculation}
\label{sec:nnlocat}

Since we perform a double expansion in $\alpha$ and $\delta$,
the NNLO computation must include all terms suppressed with
respect to the LO result by either  $\alpha^2, \ \delta^2$ or 
$\alpha \delta$. 
In order to facilitate the classification of terms we also distinguish
between radiative corrections due to hard (h),
collinear (c), and soft (s) loops, denoting the 
coupling $\alpha$ with the corresponding subscript. For example, 
a two-loop matrix element correction in the effective theory 
with no further suppression by $\delta$ will be denoted as 
$\alpha_s^2$, if both loops are soft, and if the vertices do not include 
hard corrections. With this in mind we proceed to the discussion 
of various NNLO contributions.

\subsubsection*{\it 1) Tree amplitude in the effective theory with 
insertions of LO operators in $\delta$, matched to NNLO ($\alpha_h^2$)}

The effective theory diagrams are the same as in
Figure~\ref{fig:nlohard}, but now the propagator and hard vertex 
corrections must be included to higher accuracy in $\alpha$. 
This implies the computation of $\Delta^{(3)}$, which, 
from (\ref{eq:Deltadef}) and (\ref{eq:polepos}), can be seen 
to involve $ \Pi^{(3,0)}$, the three-loop self-energy at $P^2=\hM^2$, 
as the most difficult part. In addition, the two-loop hard correction 
$C^{(2)}$ to the production (decay) vertex $\phi_v\bar\psi_{c1}
\chi_{c2}$ is required. These are standard loop calculations, 
which are difficult, but not impossible with present Feynman 
diagram technologies. Both matching coefficients,
$\Delta^{(3)}$ and $C^{(2)}$, are gauge invariant.

\subsubsection*{\it 2) Tree amplitude in the effective theory with 
insertions of NLO operators in $\delta$, matched to NLO ($\alpha_h\delta$)}

The number of possible terms of this type is limited 
by the requirement that they contribute to the tree $2\to2$
scattering amplitude in the effective theory. NLO operators can be 
constructed from the LO operators by acting on 
soft and collinear fields with extra derivatives. 
The Wilson coefficients of these operators 
then have to be matched up to NLO. Operators with bilinear 
terms in the fields require no work, since they are determined 
by the dispersion relation for the free field. Operators 
generated by adding derivatives to the LO production (decay) operator  
can be eliminated by the 
equation of motions or do not contribute at tree level for 
external states with vanishing transverse momentum. 

This leaves genuine NLO operators. An example is the four-fermion 
production-decay operator (\ref{pdop}), 
$$
 \frac{yy^*}{4 \hM^2} \,
 (\bar\psi_{c1}\chi_{c2})
 (\bar\chi_{c2}\psi_{c1}),
$$
that we encountered already in the computation of the NLO 
scattering amplitude. At NNLO we therefore need the hard 
radiative correction to the coefficient function. The corresponding 
calculation is done explicitly in the subsequent 
subsection. In general, production (decay) operators need not be 
local. Consider a collinear electron fluctuation with 
momentum $\hM n_-/2+k_1$ (where $\np k_1$ is of order $\hM$ for a 
collinear fluctuation) scattering on the neutrino with momentum 
$\hM n_+/2+k_2$, where $k_2$ is small. In this case, the scalar 
propagator is $i/(\hM n_+ k_1)$, which is far off-shell 
but does not correspond to a local interaction vertex. However, as 
discussed before, these non-local operators do not have tree matrix 
elements with the external collinear modes in 
$2\to2$ scattering.

Finally, there is a one-loop correction to the four-fermion 
production-decay operator that arises indirectly. Loop corrections 
induce an effective soft photon-neutrino coupling 
\begin{equation}
\label{nucoup}
   \frac{g a_y}{\hM^2} \,(\bar\chi_{c2}\gamma_{\nu}\chi_{c2})\, 
   \partial_\mu F_{s}^{\mu\nu} 
\end{equation}
Although this operator scales as $g\alpha_y\delta^4$, which is 
$g\alpha_y\delta^2$ suppressed relative to the neutrino kinetic term, 
it contributes to the NNLO forward scattering amplitude through 
soft-photon exchange in the $t$-channel, because the soft photon propagator 
cancels the $\delta^2$ suppression. This can be seen directly 
by converting (\ref{nucoup}) to the four-fermion operator 
using the soft photon equation of motion. In 
the following subsection we compute the coefficient function 
and discuss how this operator contributes
to the tree scattering amplitude in the effective 
theory. There exists also a similar electron-photon coupling. 
However, in this case there is no tree 
scattering diagram due to the absence of a tree-level 
neutrino-photon coupling.    

\subsubsection*{\it 3) Tree amplitude in the effective theory with 
insertions of NNLO operators in $\delta$, matched to LO ($\delta^2$)}

The task of writing down all relevant $\delta^2$ suppressed 
operators is simplified by the fact that we do not consider 
further suppressions by $\alpha$. Hence we only need to expand 
the tree $2\to 2$ scattering diagram in the underlying theory 
one order further than in (\ref{treeexpand}). We then find 
the NNLO production (decay) vertex $\bar\psi_{c1}\chi_{c2} 
(i D_{s\top})^2\phi_v$ (which in fact gives no contribution 
in the center-of-mass frame of the collision, where $k_\top=0$), 
and the NNLO production-decay operator
\begin{equation}
 -\frac{yy^*}{8 \hM^3} \,
 (\bar\psi_{c1}\chi_{c2}) \,i v D_s \,
 (\bar\chi_{c2}\psi_{c1}),
\end{equation}
which produces a correction proportional to $(\sqrt{s}-\hM)/\hM^3$ 
to the line shape.

\subsubsection*{\it 4) One-loop amplitude in the effective theory with 
insertions of LO operators in $\delta$, matched to NLO 
($\alpha_s\alpha_h,\alpha_c\alpha_h$)}

The one-loop scattering diagrams are those shown in
Figure~\ref{fig:nlosoft}, however, now there is an additional 
insertion of  $[\Delta^{(1)}]^2/4$ or $-\hM\Delta^{(2)}$ 
into the scalar propagator, or the production (decay) vertex 
is taken at NLO, involving the correction $C^{(1)}$, similar to 
the tree diagrams in Figure~\ref{fig:nlohard}. These terms 
can be obtained from the calculation in the previous section 
without further work. We simply multiply (\ref{eq:T1s}) by $C^2$,  
interpret ${\cal D}$ as $\sqrt{s}-(\hM^2+\hM(\Delta^{(1)}+
\Delta^{(2)}))^{1/2}$, expand in $\alpha$ and $\delta$ and pick out 
the NNLO terms. Note that in the underlying theory 
the two-loop diagrams corresponding to these contributions 
involve soft and hard loops, hence in the conventional terminology 
these are neither purely factorizable nor purely  
non-factorizable corrections. Nevertheless, the effective 
theory formalism guarantees that this set of terms 
is gauge-independent. 
Since the collinear one-loop contribution shown in 
Figure~\ref{fig:nlocol} vanishes, it follows that there are  
no terms proportional to $\alpha_c\alpha_h$.

\subsubsection*{\it 5) One-loop amplitude in the effective theory with 
insertions of NLO operators in $\delta$, matched to LO 
($\alpha_s\delta,\alpha_c\delta$)}

We have already discussed possible NLO operators
in the category {\it 2)}, but some of the simplifications 
based on the properties of the external momenta of the 
tree amplitude in the effective theory no longer apply to 
the one-loop amplitude. For instance, the residual momentum 
$k_\top$ of the internal scalar line no longer vanishes, and 
a non-zero contribution from the insertion of 
$(i D_{s\top})^2$ into the scalar propagator is obtained. 
In principle one may also need new production (decay) 
operators with collinear or soft photon fields in addition 
to the basic structure $\phi_v\bar\psi_{c1}\chi_{c2}$. The 
problem of deriving the operators at tree-level is very 
similar to deriving the power-suppressed heavy-to-light decay 
current in soft-collinear effective theory \cite{BCDF}. 
Since the collinear one-loop diagrams vanish, the interesting 
operators are those with a soft photon field, but it turns 
out that after applying the field equations, there is no 
gauge-invariant operator of this sort. It follows that the set of 
possible NNLO terms is again rather limited. One $\alpha_s\delta$ 
term that appears is the soft one-loop correction to the 
local four-fermion interaction shown in (\ref{eq:fourferm}). 

\subsubsection*{\it 6) Two-loop amplitude in the effective theory with 
insertions of LO operators in $\delta$, matched to LO 
($\alpha_s^2,\alpha_s\alpha_c,\alpha_c^2$)}

\begin{figure}[t]
\bigskip
\begin{center}
\begin{picture}(150,80)(0,0) 
  \ArrowLine(20,40)(10,10)
  \ArrowLine(10,70)(20,40)
  \GCirc(20,40){2}{0}
  \GCirc(80,40){2}{0}
  \Line(20,39)(80,39)
  \Line(20,41)(80,41)
  \PhotonArc(15.667,53)(12,-71,107){2}{5.5}
  \PhotonArc(50,8)(70,56.5,123.5){2}{11.5}
  \Text(60,70)[c]{\small $s$}
  \Text(35,55)[c]{\small $c$}
  \ArrowLine(90,10)(80,40)
  \ArrowLine(80,40)(90,70)
\end{picture}
\end{center}
\vskip-0.2cm
\centerline{\parbox{14cm}{\caption{\label{fig:nnlocol}\small  
Contribution to the  next-to-next-to-leading order amplitude involving
collinear (and soft) photons. }}}\end{figure}
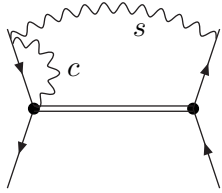

The final set of terms involves a two-loop calculation of the 
forward scattering amplitude in the effective theory, including 
vertex- and box-type corrections. However, since the propagators 
and vertices of the leading-order effective Lagrangian are 
much simpler than those in the underlying theory, the two-loop 
calculation is also greatly simplified. It is worth noting that 
in this calculation there appear for the first time non-vanishing 
collinear loop integrals. A representative diagram is 
shown in Figure~\ref{fig:nnlocol}. The requirement is that 
the external electron first emits a soft photon, so that the 
subsequent collinear emission occurs from an off-shell line. The collinear 
loop integral is then no longer scaleless.  
Because of this requirement, however, there are no contributions 
where both loops are collinear. 

\vskip0.2cm
This concludes the classification of terms. It is evident that 
a NNLO calculation of the line shape can be performed in the 
effective field theory approach without conceptual difficulties. 
In fact, the previous discussion shows that the most difficult 
calculations are reduced to standard calculations of a hard two-loop 
vertex correction and the on-shell three-loop self-energy. 
In particular, the difficulty of maintaining gauge invariance 
is absent in this approach. Since the separation of terms 
by momentum scales is gauge-independent, the sum of all contributions 
with a given power
of $\delta$, $\alpha_h$, $\alpha_c$, and $\alpha_s$ is 
separately gauge-independent.

In the version of the 
effective theory where collinear fluctuations are integrated 
out, the collinear loop contributions move from the 
effective theory matrix element to a matching correction. 
The calculations to be performed remain of course the same, 
but the structure of operators is different, since 
the effective theory no longer contains fields for collinear 
fluctuations. While this simplifies the field content 
of possible operators, it also complicates their structure, 
introducing new non-localities related to the fact that 
one component of collinear momentum is of the same order 
as the soft momentum. There is again a close analogy with 
recent developments in soft-collinear effective theory, 
which, however, we do not pursue further here. 

\subsection{One-loop matching of the production-decay operator}
\label{sec:pd4}

In this subsection we perform an explicit calculation of the 
$\alpha_h\delta$ contributions to the 
forward scattering amplitude categorized under {\it 2)} 
above.  

\subsubsection{Four-fermion operator}

The four-fermion operator (\ref{pdop}) is a sub-leading operator that 
gives rise to a production-decay vertex. It contributes to the 
forward scattering amplitude at NLO through the tree 
diagram (\ref{eq:fourferm}). We now determine the loop correction to the  
coefficient function of this vertex. 

\begin{figure}[t]
\vskip0.1cm
\begin{center}
\begin{picture}(300,80)(0,0) 
  \ArrowLine(20,40)(10,10)
  \ArrowLine(10,70)(20,40)
  \DashLine(20,40)(80,40){4}
  \PhotonArc(50,40)(15,0,180){2}{7}
  \ArrowLine(90,10)(80,40)
  \ArrowLine(80,40)(90,70)
  \SetOffset(100,0)
  \ArrowLine(20,40)(10,10)
  \ArrowLine(10,70)(20,40)
  \DashLine(20,40)(80,40){4}
  \PhotonArc(25,41)(23,-2,120.5){2}{7}
  \ArrowLine(90,10)(80,40)
  \ArrowLine(80,40)(90,70)
  \SetOffset(200,0)
  \ArrowLine(20,40)(10,10)
  \ArrowLine(10,70)(20,40)
  \DashLine(20,40)(80,40){4}
  \PhotonArc(50,10)(60,53,127){2}{9.5}
  \ArrowLine(90,10)(80,40)
  \ArrowLine(80,40)(90,70)
\end{picture}
\end{center}
\vskip-0.3cm
\centerline{\parbox{14cm}{\caption{\label{fig:box}\small  
Diagrams in the full theory proportional to $\alpha_y\alpha_g$, 
whose hard part contributes to
the matching coefficient of the production-decay vertex at NLO. A
diagram with a vertex correction at the decay vertex is
understood. }}}
\end{figure}
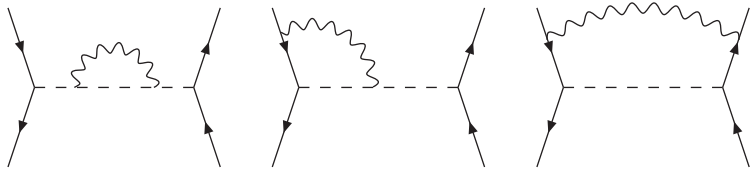

Consider the box diagram contribution to the scattering amplitude 
in the full theory shown in Figure~\ref{fig:box}. It is clear that the
hard part of this diagram results in a loop correction to a 
local four-fermion operator. This contribution alone is gauge-dependent. 
However, there are additional contributions from vertex and self-energy
corrections (also shown in Figure~\ref{fig:box}), since 
expanding the corresponding subgraphs in $\delta$ results in 
terms where the propagators of the unstable scalar are canceled. 
Because of this cancellation these terms also contribute to the 
coefficient of the local four-fermion operator, rendering the  
complete result gauge-independent. 

To see this, consider first the terms proportional 
to $\alpha_y\alpha_g$ in Figure~\ref{fig:box}. Due to the two
scalar propagators, the leading contribution from the 
self-energy diagram is of order
$\alpha^2/\delta^2$. This is a LO contribution, which is
already included in the effective theory through the propagator 
$i/(2\hM (vk)-\hM \Delta^{(1)})$ of
the scalar field. The next term in the expansion in $\delta$ 
cancels one propagator, resulting in a local vertex. This is 
already included in the computation of the matching coefficient
$C^{(1)}$. The second term in the expansion in $\delta$, 
however, results in a NNLO contribution, which has to be reproduced by
the effective theory via the matching coefficient of the
production-decay vertex. Similarly, the leading contribution 
from the vertex diagram is included in $C^{(1)}$, and we need 
the next term in the expansion in $\delta$, where the 
scalar propagator is canceled. We then obtain for 
the contribution from the self-energy, vertex and box diagrams
\beq
B^{(2)}_{\rm se} &=& 
i |y|^2 \frac{[\bar{u}(p)v(q)]\,[\bar{v}(q)u(p)]}{\hM^2} \,
\Pi^{(1,2)} \nonumber\\  
&=& \frac{i |y|^2 a_g}{\hM^2} [\bar{u}(p)v(q)]\,[\bar{v}(q)u(p)] \,
\left(-\frac{3-\xi}{2\,\epsilon} + \frac{3-\xi}{2} \lnM 
+ (1-\xi) \right),
\label{eq:Bcont1}\\[0.2cm]
B^{(2)}_{\rm v} &=&
\frac{i |y|^2 a_g}{\hM^2} [\bar{u}(p)v(q)]\,[\bar{v}(q)u(p)]\,
\left(- \frac{2}{\epsilon^2} + \frac{2-\xi}{\epsilon} 
+ \frac{2}{\epsilon} \lnM  \right. \nonumber\\
&& \hspace*{0cm}\left. -\  (2-\xi) \lnM - \lnsqM
- 2 (4-\xi) - \frac{\pi^2}{6} \right),
\label{eq:Bcont2}\\[0.2cm]
B^{(2)}_{\rm b} &=&
\frac{i |y|^2 a_g}{\hM^2} [\bar{u}(p)v(q)]\,[\bar{v}(q)u(p)]
\,\left(-\frac{1-\xi}{2 \epsilon} + \frac{1-\xi}{2} \lnM +(1-\xi) \right)
\nonumber\\
&+&
\frac{i |y|^2 a_g}{\hM^2} [\bar{u}(p) \gamma^\mu \gamma^\nu v(q)]\,
[\bar{v}(q)\gamma_\nu \gamma_\mu u(p)] \,
\left(-\frac{1}{2\epsilon} + \frac{1}{2} \lnM -
\frac{3}{2} \right).
\label{eq:Bcont3}
\eeq
Summing (\ref{eq:Bcont1}), (\ref{eq:Bcont2}) and
(\ref{eq:Bcont3}) we see that the gauge dependence cancels, and we
obtain 
\be
\frac{i B_1}{\hM^2} [\bar{u}(p)v(q)]\,[\bar{v}(q)u(p)] +
\frac{i B_2}{\hM^2} [\bar{u}(p) \gamma^\mu \gamma^\nu v(q)]\,
[\bar{v}(q)\gamma_\nu \gamma_\mu u(p)]
\label{eq:ff}
\ee
with
\beq
B_1 &=& - |y|^2 a_g \left( 
  \frac{2}{\eps^2} - \frac{2}{\eps} \lnM + \lnsqM + 6 +
  \frac{\pi^2}{6} \right),  \nonumber \\
B_2 &=& - |y|^2 a_g \left(\frac{1}{2 \eps} 
- \frac{1}{2} \lnM + \frac{3}{2} \right).
\eeq
The cancellation of the gauge dependence is of course not
accidental. Since the separation of hard and soft parts is 
gauge-invariant, and since we match $S$-matrix elements, the 
short-distance coefficient must be gauge-independent. 

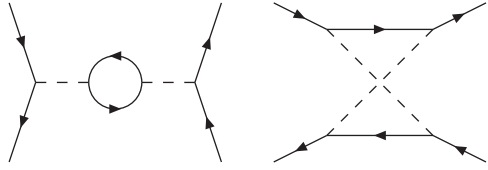
\begin{figure}[t]
\vskip0.1cm
\begin{center}
\begin{picture}(200,80)(0,0) 
  \ArrowLine(20,40)(10,10)
  \ArrowLine(10,70)(20,40)
  \DashLine(20,40)(40,40){4}
  \DashLine(60,40)(80,40){4}
  \ArrowArc(50,40)(10,0,180)
  \ArrowArc(50,40)(10,180,0)
  \ArrowLine(90,10)(80,40)
  \ArrowLine(80,40)(90,70)
  \SetOffset(100,0)
  \ArrowLine(30,20)(10,10)
  \ArrowLine(70,20)(30,20)
  \ArrowLine(90,10)(70,20)
  \DashLine(30,20)(70,60){4}
  \DashLine(70,20)(30,60){4}
  \ArrowLine(70,60)(90,70)
  \ArrowLine(30,60)(70,60) 
  \ArrowLine(10,70)(30,60)
\end{picture}
\end{center}
\vskip-0.3cm
\centerline{\parbox{14cm}{\caption{\label{fig:NNLO/hard/phi}\small  
Diagrams in the full theory proportional to $\alpha_y^2$,
whose hard part contributes to
the matching coefficient of the production-decay vertex at NLO.}}}

\end{figure}

There are further contributions to the matching coefficient. 
First, the terms proportional to $\alpha_y^2$ from the diagrams 
shown in Figure~\ref{fig:NNLO/hard/phi} are given by
\begin{equation}
\frac{i B_3}{\hM^2} [\bar{u}(p)v(q)]\,[\bar{v}(q)u(p)] +
\frac{i B_4}{\hM^2} [\bar{u}(p) \slash{q} u(p)]\,
[\bar{v}(q)\slash{p} v(q)]+
\frac{i B_5}{\hM^2} [\bar{u}(p) \gamma^\mu u(p)]\,
[\bar{v}(q) \gamma_\mu v(q)]
\label{eq:ff/phi}
\end{equation}
with
\be
B_3 = - |y|^2 a_y, 
\qquad
B_4 = - |y|^2 a_y \left( 
  -5+\frac{\pi^2}{2}\right), 
\qquad
B_5 = - |y|^2 a_y \left( 
  1-\frac{\pi^2}{12}
  \right). 
\ee
Second, up to now we have been discussing contributions which 
(with the exception of the box diagrams) come from expanding 
the one-particle irreducible self-energy or vertex subgraphs in 
$\delta$. Additional contributions arise from the expansion 
of the one-particle reducible scalar propagator according to 
(\ref{treeexpand}). For 
instance, combining the local NLO term from (\ref{treeexpand}) 
with the (renormalized) hard vertex at leading order in $\delta$, we 
obtain a contribution to the coefficient of the 
four-fermion operator proportional to $C^{(1)}$. Similar terms 
come from the self-energy diagram. 
Collecting all contributions, we obtain for the NLO correction 
to the production-decay operator (\ref{pdop})
\begin{eqnarray}
T^{(1)} &=&  -\frac{|y|^2}{\hM^2} \,
 (\bar{\psi}_{c1}\chi_{c2})(\bar{\chi}_{c2}\psi_{c1}) 
 \Bigg[a_y + a_g \left( \frac{2}{\eps^2} - \frac{2}{\eps} \lnM + \lnsqM +
      \frac{\pi^2}{6} + 6 \right) 
\nonumber\\
&&  \hspace*{1cm} -\frac{C^{(1)}}{2} + \frac{\Delta^{(1)}}{16\hM} \Bigg]
\nonumber\\[0.2cm]
 &&  - \frac{|y|^2}{\hM^2}\,
    \big(\bar{\psi}_{c1}\gamma_\nu\gamma_\mu\chi_{c2}\big) 
    \left(\bar{\chi}_{c2}\gamma^\mu\gamma^\nu\psi_{c1}\right)  
    \, a_g \left(\frac{1}{2\eps}-\frac{1}{2}\lnM + \frac{3}{2}\right) \
\nonumber\\[0.2cm]
&&    - \frac{|y|^2}{\hM^2} \,
    \Big(\bar{\psi}_{c1} \frac{\slash{n}_+}{2} \psi_{c1}\Big) 
    \Big(\bar{\chi}_{c2} \frac{\slash{n}_-}{2} \chi_{c2}\Big)
    \, a_y \left(-5+\frac{\pi^2}{2}\right) 
\nonumber\\[0.2cm] 
&&   - \frac{|y|^2}{\hM^2}\, 
    \big(\bar{\psi}_{c1} \gamma^{\mu} \psi_{c1}\big) 
    \left(\bar{\chi}_{c2} \gamma_{\mu} \chi_{c2}\right)
    \,  a_y \left(1-\frac{\pi^2}{12}\right), 
\label{tnlo}
\end{eqnarray}
with $\Delta^{(1)}/\hM$ and $C^{(1)}$ given in (\ref{eq:Delta1}) 
and (\ref{C1final}), respectively. Inserting the explicit expressions for 
these quantities, the poles in $\epsilon$ do not cancel. The 
left-over poles are related to the factorization of hard, collinear 
and soft contributions, and the initial-state collinear 
singularity, and are canceled only once all other 
NNLO contributions to the scattering amplitude are combined. 
The final result for the matching coefficient is obtained from 
the above expression with all poles subtracted minimally. 

We expressed (\ref{tnlo}) in terms of four-fermion operators with 
four Dirac structures as they come out of the calculation. Not 
all of them are independent. Using Fierz transformations and
the projection properties $\slash{n}_-\psi_{c1}=
\slash{n}_+\chi_{c2}=0$ of the collinear fermion fields, they can be reduced 
to linear combinations of the three basis operators 
\be
(\bar{\psi}_{c1}\chi_{c2})(\bar{\chi}_{c2}\psi_{c1}), \qquad 
(\bar{\psi}_{c1}\gamma_5\chi_{c2}) (\bar{\chi}_{c2}\gamma_5\psi_{c1}), \qquad 
(\bar{\psi}_{c1}\gamma_{\mu_\perp}\chi_{c2})
(\bar{\chi}_{c2}\gamma^{\mu_\perp}\psi_{c1}),
\ee 
where ``$\mu_\perp$'' means that one sums only over the transverse 
components. To perform this reduction, it is necessary to define 
carefully the Dirac algebra in $d$ dimensions in very much the same 
way as in applications of the weak effective Hamiltonian 
to weak interaction processes at low energies.

\subsubsection{Effective photon-neutrino coupling}

\begin{figure}[t]
\vskip0.1cm
\begin{center}
\begin{picture}(200,80)(0,0) 
  \ArrowLine(50,60)(90,70) 
  \ArrowLine(10,70)(50,60)
   \ArrowLine(30,20)(10,10)
   \DashLine(30,20)(50,40){3}
   \DashLine(50,40)(70,20){3}
   \ArrowLine(70,20)(30,20)
   \ArrowLine(90,10)(70,20)
   \Photon(50,40)(50,60){3}{3}
  \SetOffset(100,0)
  \ArrowLine(50,60)(90,70)
  \ArrowLine(10,70)(50,60)
   \ArrowLine(30,20)(10,10)
   \ArrowLine(50,40)(30,20)
   \ArrowLine(70,20)(50,40)
   \DashLine(30,20)(70,20){3}
   \ArrowLine(90,10)(70,20)
   \Photon(50,40)(50,60){3}{3}
\end{picture}
\end{center}
\vskip-0.3cm
\centerline{\parbox{14cm}{\caption{\label{fig:chichiF}\small  
NNLO forward scattering diagrams in the full theory with 
a soft photon exchanged in the $t$-channel. The hard contribution 
to the triangle subgraph generates the effective photon-neutrino 
coupling $(\bar{\chi}\chi) \partial F$.
}}}
\end{figure}
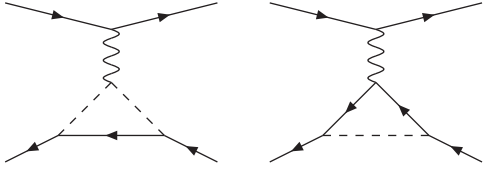

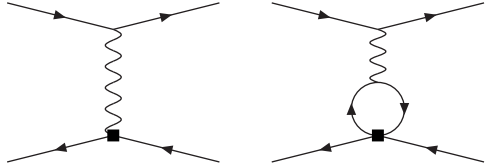
\begin{figure}[t]
\vskip0.1cm
\begin{center}
\begin{picture}(200,80)(0,0)
  \ArrowLine(50,20)(10,10)
  \ArrowLine(90,10)(50,20)
  \ArrowLine(50,60)(90,70) 
  \ArrowLine(10,70)(50,60)
  \Photon(50,20)(50,60){3}{5} 
  \GBox(48,18)(52,22){0}
  \SetOffset(100,0) 
  \ArrowLine(50,20)(10,10)
  \ArrowLine(90,10)(50,20)
  \ArrowLine(50,60)(90,70)
  \ArrowLine(10,70)(50,60)
  \Photon(50,40)(50,60){3}{3} 
  \ArrowArcn(50,30)(10,90,-90)
  \ArrowArcn(50,30)(10,-90,90)
  \GBox(48,18)(52,22){0}
\end{picture}
\end{center}
\vskip-0.3cm
\centerline{\parbox{14cm}{\caption{\label{fig:chichiF/eff}\small  
Effective theory tree diagrams related to the full theory 
diagrams in Figure~\ref{fig:chichiF}. The left-hand diagram involves  
the $(\bar{\chi}\chi) \partial F$ coupling, the right-hand diagram 
the mixed collinear-soft four-fermion operator and a 
soft electron loop.}}}
\label{fig:NNLO/hard/t-channel}
\end{figure}

There is a second correction of order $\alpha_h\delta$, which we discuss 
separately, although it can be interpreted as a further contribution 
to (\ref{tnlo}). Consider the two diagrams in the full theory displayed 
in Figure~\ref{fig:chichiF}, which do not involve resonant scalar 
lines. The hard contribution to the triangle subgraphs (consisting 
of scalar and electron lines) induces an effective 
soft-photon neutrino coupling.  
The result of the calculation gives
\be
\label{ncoup}
  \frac{g a_y}{\hM^2} 
  \left(\bar{\chi}_{c2}\gamma_{\nu}\chi_{c2}\right)
  \partial_\mu F_{s}^{\mu\nu}
  \left[
    \frac{1}{3\eps} -\frac{1}{3}\ln\frac{\hM^2}{\mu^2}
    +\frac{1}{2}
  \right].
\ee 
In Figure~\ref{fig:chichiF} the photon must be soft, because 
we consider the forward scattering amplitude. (In fact, to regulate 
the otherwise divergent $t$-channel propagator, we must temporarily 
consider the amplitude slightly non-forward.) Power counting shows that 
this is a $\delta^2$ suppressed operator. The factor of $\delta^2$ 
is compensated by the $t$-channel photon propagator of order 
$1/\delta^2$, the result being an $\alpha_g\alpha_y$ (NNLO) 
contribution to the scattering amplitude in the 
effective theory shown as the left-hand 
diagram of Figure~\ref{fig:chichiF/eff}. The photon propagator is 
canceled by a momentum factor from the vertex (\ref{ncoup}) 
resulting in a local contribution to the scattering amplitude of the 
same form as (\ref{tnlo}). This can be seen explicitly using the 
soft photon equation of motion 
$\partial_\mu F_{s}^{\mu\nu}=-g \,\bar\psi_{c1}
\gamma^\nu\psi_{c1} + \ldots$ to rewrite (\ref{ncoup}) 
as a four-fermion operator times a coefficient proportional to 
$|y|^2 a_g$. 

The $1/\epsilon$ pole in (\ref{ncoup}) is a soft divergence, 
which arises due to the hard-soft factorization from the second 
diagram in Figure~\ref{fig:chichiF}. 
The divergence cancels with the soft contribution to this 
diagram, where the two internal electron lines are soft, and 
the scalar line is off-shell. The off-shell scalar induces 
a local $(\bar{\chi}_{c2}\psi_s) (\bar{\psi}_s \chi_{c2})$ vertex. 
This mixed collinear-soft four-fermion operator contributes 
to the forward scattering amplitude in the effective 
theory through the one-loop diagram shown on the 
right-hand side of Figure~\ref{fig:chichiF/eff}, where the 
electron loop is soft. Once again 
the $\delta^2$ suppression of the operator is compensated 
by the soft photon propagator. 

\section{Summary and outlook}
\label{sec:conclude}

It has been recognized for some time that the perturbative treatment 
of unstable particles is difficult, because partial summations of the 
perturbative expansion are necessary. These difficulties originated 
mainly from the fact that the guiding principle for resummation was 
not understood. In this paper we advocated the idea 
that effective field theory methods familiar from other applications 
in high-energy physics can be adapted to solve the problem of describing 
systematically the production and decay of unstable particles, since  
the fundamental reason for the breakdown of weak-coupling perturbation theory 
is related to the emergence of a second (small) momentum scale 
near resonance. 
The advantages of the effective theory method are:
\begin{itemize}
\item It breaks the calculation into several well-defined pieces 
(matching calculations, matrix element calculations), thus rendering 
the organization of the calculation efficient and transparent.
\item It provides a power-counting scheme in the small parameters 
($\delta$, couplings), which allows for an identification of the terms 
relevant for achieving a prescribed accuracy before actual calculations 
must be done.
\item It provides a set of (Feynman) rules to compute the minimal 
set of terms necessary for a given accuracy. Since one does not 
calculate ``too much'', the calculation to a given order is presumably 
technically simpler than in any other approach.
\item Gauge invariance is automatic at every order, since the 
effective Lagrangian is gauge-invariant. 
\item It can be extended to any accuracy in the expansion in 
$\delta$ and in couplings at the expense of performing more complicated, 
but well-defined calculations.
\end{itemize}
In this paper we explained these ideas in a toy theory restricting 
ourselves to a single scalar resonance, an abelian gauge theory, 
and a totally inclusive resonant scattering cross section. Extending 
the method to non-abelian gauge theories or relaxing 
any of the other restrictions causes only technical complications. For 
instance, for a resonant fermion, the heavy scalar Lagrangian 
would be replaced by the familiar heavy quark effective Lagrangian. 
For resonant gauge bosons in the electroweak Standard Model in 
$R_\xi$ gauge with $\xi$ not near one, the degrees of freedom with 
mass $\xi \hM^2$ are integrated out, and the effective theory 
contains a massive vector field with three polarizations. 
Non-renormalizability is not an issue, since the effective field 
describes only soft fluctuations. In 't~Hooft-Feynman gauge, 
or with $\xi$ close to one, the effective theory contains a vector 
field with four polarizations and a pseudo-Goldstone field, which 
cancels the effect of the scalar polarization state, so 
that one can work again with a canonical massive vector field. 

Other obvious extensions of this work concern non-inclusive 
kinematics and pair production of unstable particles. In these cases 
it will be necessary to enlarge the field content of the 
effective theory. For instance, if energetic particles (jets) are 
detected in the final state, one must introduce new collinear fields 
corresponding to the direction of these particles. It is advantageous 
to work with cut diagrams rather than amplitudes, so that 
the (final-state weighted) phase-space integrals can be treated 
on the same footing as loop diagrams in matching calculations. 
Similarly the extension to pair production requires the introduction 
of two copies of what we called the HSET Lagrangian in this 
paper, one for each unstable particle with different velocity 
vectors. The production (decay) vertices then depend on additional 
kinematic quantities such as the scalar product of the velocities. 
Pair production near threshold appears in this context as the 
particular case, where the two velocities become nearly 
equal. The two HSET Lagrangian then merge to the non-relativistic 
Lagrangian, and standard methods can be applied to deal 
with the potential complications due to the strong Coulomb force. 
We are confident that effective field theory will be the 
method of choice to address unstable particle production in 
these different kinematic situations and plan to return 
to some of them in the context of concrete applications.

\subsubsection*{Acknowledgements}

We thank M.~Kalmykov for correspondence. 
The work of M.B. and A.C. is supported by the DFG 
Sonder\-forschungs\-bereich/Trans\-regio 9 
``Computer-gest\"utzte Theoretische Teilchenphysik''. We 
also acknowledge the use of the Mathematica packages FeynCalc and 
FeynArts \cite{feyncalc/feynarts} for performing some of the 
loop calculations.  

\section*{Appendix}
\label{sec:appendix}

In this appendix we collect the renormalization factors and
counterterms of the theory defined by the Lagrangian (\ref{model}). 
The counterterm Lagrangian is
\begin{eqnarray}
{\cal L}_{\rm ct} &=& 
\delta_\phi \,\partial_\mu\phi^\dagger \partial^\mu\phi - 
\delta_M \hat M^2\phi^\dagger\phi + 
\delta_\psi \bar\psi i\slash{\partial}\psi + 
\delta_\chi \bar\chi i\slash{\partial}\chi -
\frac{\delta_A}{4} F^{\mu\nu}F_{\mu\nu}
\nonumber\\
&&+\,\delta_g g \mu^\epsilon \bar\psi \slash{A}\psi+
\delta_g^\prime g\mu^\epsilon \left(\phi^\dagger A_\mu i\partial^\mu\phi-
(i\partial^\mu\phi^\dagger) A_\mu\phi\right)+
\delta_{g^2} g^2 \mu^{2\epsilon} \phi^\dagger A_\mu A^\mu\phi
\nonumber\\
&&+\,\delta_y y\mu^\epsilon  \phi\bar\psi\chi+\delta_y^*y^*
\mu^\epsilon \phi^\dagger \bar\chi\psi
-\frac{\delta_\lambda}{4}\mu^{2\epsilon}(\phi^\dagger \phi)^2.
\end{eqnarray}
The relation between bare and renormalized quantities is
\begin{eqnarray}
\begin{array}{ll}
\phi_0=\sqrt{Z_\phi} \phi, & \delta_\phi = Z_\phi-1,\\
M_0^2=Z_M^2 \hat M^2, \quad & \delta_M = Z_M^2 Z_\phi-1, \qquad\\
\psi_0=\sqrt{Z_\psi} \psi, & \delta_\psi = Z_\psi-1,\\
\chi_0=\sqrt{Z_\chi} \chi, & \delta_\chi = Z_\chi-1,\\
A_0=\sqrt{Z_A} A, & \delta_A = Z_A-1,\\
\xi_0=Z_A \xi, & \\
g_0=Z_g g\mu^\epsilon , \quad & \delta_g = Z_g Z_\psi\sqrt{Z_A}-1,\\
y_0=Z_y y\mu^\epsilon , \quad & 
\delta_y = Z_y \sqrt{Z_\phi}\sqrt{Z_\psi}\sqrt{Z_\chi}-1,\\
\lambda_0=(\lambda+\Delta_\lambda) \mu^{2 \epsilon} & 
\delta_\lambda = (Z_\phi^2-1)\lambda+Z_\phi^2 \Delta_\lambda,
\end{array}
\end{eqnarray}
The counterterms $\delta_g^\prime$ and $\delta_{g^2}$ are not independent, 
but determined by the previous ones:
\begin{eqnarray}
\delta_g^\prime &=& Z_g Z_\phi\sqrt{Z_A}-1,
\nonumber\\
\delta_{g^2}^\prime &=& Z_g^2 Z_\phi Z_A-1.
\end{eqnarray}
In this paper all renormalization constants are needed at the 
one-loop order except for $Z_M^2$, which must be known to two loops. 
In the $\overline{\rm MS}$ scheme, and in covariant gauge with 
gauge parameter $\xi$, the counterterms expressed in terms of
the renormalized couplings are given by: 
\begin{eqnarray}
Z_y &=& 1+
  \frac{1}{\eps}
  \left(  - \frac{3}{2} \, a_g + \frac{3}{2} \, a_y \right) 
\label{eq:Zy} \nn \\
Z_g &=& 1+
  \frac{1}{\eps}
  \left( \frac{5}{6} \, a_g \right)
\label{eq:Zg}  \nn \\
Z_\phi &=& 1+
  \frac{1}{\eps}
  \left( (3-\xi)\, a_g - 2\, a_y  \right)
\label{eq:Zphi} \nn  \\
Z_\psi &=& 1+
  \frac{1}{\eps}
  \left( -\xi a_g - \frac{1}{2}\, a_y  \right)
\label{eq:Zpsi}  \nn \\
Z_\chi &=& 1+
  \frac{1}{\eps}
  \left( - \frac{1}{2}\, a_y \right)
\label{eq:Zchi}  \nn \\
Z_A &=& 1+
  \frac{1}{\eps}
  \left( - \frac{5}{3}\, a_g \right)
\label{eq:ZA}  \nn \\
  Z_M^{\,2} &=& 1 + 
  \frac{1}{\eps}
  \left( - 3\, a_g + 2\, a_y\right) +
  \frac{1}{\eps^2}
  \left( 8\, a_g^2 + a_y^2 - 9\, a_g a_y \right) 
  \nn \\
  && + \ \frac{1}{\eps}
  \left( \frac{53}{6}\, a_g^2 - \frac{3}{2}\, a_y^2
         + \frac{5}{2}\, a_g a_y +a_\lambda\right)
\label{eq:ZM}  
\end{eqnarray}
Furthermore, we need the part of $\delta_\lambda$ that is proportional
to $\alpha_y^2$ and $\alpha_g^2$. It is given by
\be
\delta_\lambda = -\frac{8\, \alpha_y^2}{\eps}
    +\frac{12\, \alpha_g^2}{\eps}.
\label{eq:deltaLambda}
\ee

The ultraviolet-finite residues of the $\MSbar$-renormalized 
propagators read:
\beq
R_\phi &=& 1 - (3-\xi)\, \frac{1}{\eps} a_g
   + a_y \left(2 \log \frac{\hM^2}{\mu^2}-2 - 2\, i \pi \right) 
\label{eq:Rphi} \nn \\
R_\psi &=& 1 +  \frac{\xi}{\eps} a_g
   + a_y \left(\frac{1}{2} \log \frac{\hM^2}{\mu^2}
                  - \frac{1}{4} \right) 
\label{eq:Rpsi}  \nn \\
R_\chi &=& 1+ a_y
   \left(\frac{1}{2} \log \frac{\hM^2}{\mu^2}- \frac{1}{4} \right) 
\label{eq:Rchi} \nn  \\
R_A &=& 1+ a_g
   \left(\frac{1}{3} \log \frac{\hM^2}{\mu^2} + \frac{4}{3\eps} \right) 
\label{eq:RA}
\eeq
The $1/\epsilon$ poles in these equations are infrared divergences. 
In the dimensional infrared regularization scheme the full residues 
equal their hard contributions, so $R_{hX}=R_X$.

\end{document}